\newcommand{\ie}{\emph{i.e.}}

\documentclass[conference, letterpaper]{IEEEtran}

\usepackage{multirow,array}
\ifCLASSINFOpdf
\else
\fi
\usepackage{url}

\usepackage{subcaption}

\ifCLASSINFOpdf
   \usepackage[pdftex]{graphicx}
\else
\fi

\usepackage[cmex10]{amsmath}
\usepackage{color}
\usepackage{fancyhdr}

\renewcommand{\thispagestyle}[2]{}

\fancypagestyle{plain}{
        \fancyhead{}
        \fancyhead[C]{first page center header}
        \fancyfoot{}
        \fancyfoot[C]{first page center footer}
}
\begin{document}

%
\title{An interacting replica approach applied to the traveling salesman problem}

\author{\IEEEauthorblockN{Bo Sun, Blake Leonard, Peter Ronhovde \\ and Zohar Nussinov*}
\IEEEauthorblockA{Department of Physics\\
Washington University in St. Louis\\
Saint Louis, Missouri 63130, USA\\
Email: zohar@wuphys.wustl.edu}

}


%


\maketitle

\begin{abstract}
We present a physics inspired heuristic method for solving combinatorial optimization problems. Our approach is specifically motivated by the desire to avoid trapping in metastable local minima- a common occurrence in hard problems with multiple extrema. Our method involves (i) coupling otherwise independent simulations of a system (``replicas'') via geometrical distances as well as (ii) probabilistic inference applied to the solutions found by individual replicas. The {\it ensemble} of replicas evolves as to maximize the inter-replica correlation while simultaneously minimize the local intra-replica cost function (e.g., the total path length in the Traveling Salesman Problem within each replica). We demonstrate how our method improves the performance of rudimentary local optimization schemes long applied to the NP hard Traveling Salesman Problem. In particular, we apply our method to the well-known ``$k$-opt''  algorithm and examine two particular cases- $k=2$ and $k=3$. With the aid of  geometrical coupling alone, we are able to determine for the optimum tour length on systems up to $280$ cities (an order of magnitude larger than the largest systems typically solved by the bare $k=3$ opt).  The probabilistic replica-based inference approach improves $k-opt$ even further and determines the optimal solution of a problem with $318$ cities and find tours whose total length is close to that of the optimal solutions for other systems with a larger number of cities.

\end{abstract}



%
\IEEEpeerreviewmaketitle

\section{Introduction}  \label{sec:introduction}
Combinatorial optimization problems \cite{ref:papadimitriou1998} often possess
a relatively large number of locally optimal pseudo-solutions, similar to the
abundance of metastable energy states in complex physical systems.
This can make determination of the global optimum difficult, especially for heuristic
algorithms which attempt to optimize a cost function locally (\ie{}, by iteratively
perturbing the parameters, testing the resulting change in the cost function, 
and allowing the state change if the cost function is decreased or some other
conditions are satisfied).

The concept of a local optimizer acting on some cost function in parameter space
is equivalent to modeling a thermal system exploring its energy landscape.
At a certain temperature, a thermal system can realistically exchange a certain amount
of heat with the environment.  While the system is generally attempting to find the
lowest energy state, it can temporarily gain energy, and, in so doing escape from a
local energy well.
However, if the well is deeper than the realistically allowable energy gain, then the
system may remain stuck in a metastable, locally optimal energy state indefinitely.
This is what happens in spin glasses \cite{ref:mezard1987}, for instance.

Analogously, if a local well of the cost function in parameter space is deeper than
the realistically allowable positive gain in the cost function, then the simulation
of a local optimizer will remain stuck, creating a design tradeoff.
The greater potential gain allowed in the cost function, the easier it is for the
simulation to escape potential wells, but the longer it will take to actually find
a minimum because it will have a larger search space at each step, and it will be
moving ``uphill'' more often.  This is the reason that heuristic algorithms can generally
be relied upon to produce good pseudo-solutions a few percent above the optimum value,
but rarely find the actual global optimum in sufficiently complex problems.

Previous methods such as replica exchange \cite{ref:percus2008} and genetic algorithms 
\cite{ref:g222} have attempted to address this problem with varying degrees of efficacy 
depending on context.
We were inspired to use ``information-based replica'' correlations  and inference to systematically detect ideal subgraph (or ``community'') partitions of a large graph as two of us have done several years ago  \cite{ref:peter2009}. By ``replicas'' we here allude to independent copies of the same problem. Since then these notions have been applied to a variety of complex system physics (both static and dynamic) and image segmentation problems \cite{peter1,peter2,dandan1,dandan2}. More recently, other works applied similar notions to a host of interesting problems \cite{multiplex1,multiplex2,multiplex3}. Free-energies and entropies 
of such ensembles or ``multiplexes''  have been discussed in \cite{ref:peter2009,ref:bianconi2013}.
In our approach we do not focus solely on directly extracting the minimum amongst an ensemble of solvers. A key ingredient that we introduced in earlier work is the use of inference to predict
which features of the solutions appearing in disparate replicas may coincide with those in the optimal solution;
this inference coupling as well as other effective ``interactions'' between the replicas (e.g., a ``geometrical coupling'' discussed below)
between the replicas may lead to solutions that at intermediate steps elevate the energy
(yet lower a ``free energy'' \cite{ref:peter2009}) similar to the way in which thermal effects may, at intermediate steps, 
elevate energies in annealing algorithms. Transitions in the complexity of combinatorial optimization problems such community detection problems have crisp signatures in inter-replica correlation functions and information theory measures\cite{dandan3}. Augmenting Refs. 
\cite{ref:peter2009,peter1,peter2,dandan1,dandan2,dandan3}, we further also note the more recent work  of Ref. \cite{ref:saad2013} in which the authors 
demonstrate that the inference algorithms based on evolving interactions
between replicated solutions in a cavity type approach have better performance in the binary Ising percepton problem.
We further note that the ``wisdom of the crowds" (which we took in Refs. \cite{ref:peter2009,peter1,peter2,dandan1,dandan2,dandan3} to be independent replicas) 
has been long appreciated \cite{ref:james}.

The approach that we further develop in the current work- that of geometric interactions between individual solvers and probabilistic ensemble inference- emulates the biological and sociological advantages long known colloquially from collective behaviors and ``wisdom of the crowds''  \cite{ref:james}.
Historically, biologically inspired ``swarm intelligence'' algorithms \cite{ref:clerc} have spawned algorithms such as the well known Ant Colony System (ACS) \cite{ref:dorigo1997b} with which we will later compare the new probabilistic variant of our method.
In a broad sense, the spin-glass physics cavity approximation inspired message type algorithms of the type of Ref. \cite{ref:saad2013}, exchange Monte Carlo \cite{ref:percus2008}, genetic algorithms \cite{ref:g222}, the work that two of us developed in Refs. \cite{ref:peter2009,peter1,peter2,dandan1,dandan2,dandan3}, the ACS, and a multitude of other approaches might all be cast as particular realizations of broad ensemble based interactions or moves.

In the current work, we present new optimization methods based upon the concepts of (i) geometrical distance 
coupling (GDC) and (ii)  inference amongst independent replicas. We apply these tools to the Traveling Salesman Problem. We demonstrate that while a single replica can do quite poorly
in solving a challenging problem,  via the use of (i) and (ii) above, the {\it ensemble} of replicas can address much harder problems.

\begin{figure}[h]
\centering
\includegraphics[width=0.95\columnwidth]{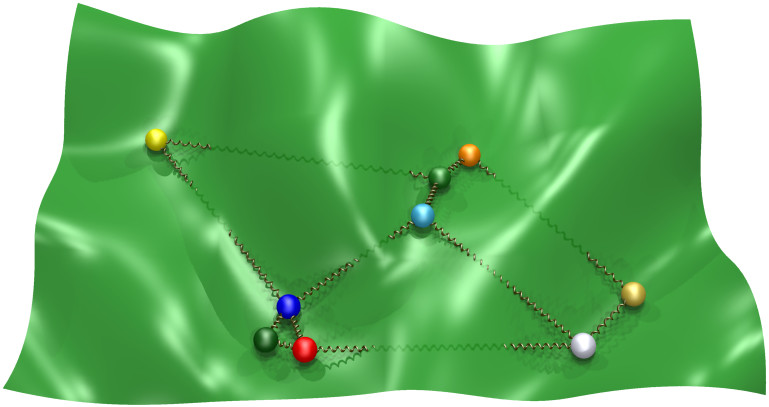}
\caption{(Color online.)
Coupled replicas in a high dimensional energy landscape.
The springs schematically represent the tendency of replicas to collectively
interact with one another when veering towards viable minima.}
\label{fig:figzero}
\end{figure}

We employ a quantity, which we term the ``GDC distance'' $C(A,B) \ge 0$ (see Appendix, Eq. (A2) in particular)
to measure the similarity between two tours A and B.
A distance C(A,B) =0 indicates that tours A and B are identical.
We coupled otherwise independent optimizers via their geometrical distances,
so that the optimizers will essentially have two ``forces'' influencing their behavior, see Fig. \ref{fig:figzero}.
Each optimizer will (i) independently desire to decrease its cost function locally,
while simultaneously attempting to (ii) minimize the GDC between itself and all other optimizers (portrayed by the harmonic springs in Fig.  \ref{fig:figzero}). 
We demonstrate that through this coupling, local optimizers can escape from wells which
otherwise would have confined them permanently, in the cases we studied.

An additional central tool that we will invoke in this work is that
of probabilstic inference from the different replicas, e.g., \cite{ref:peter2009,dandan1} or a ``replica inference based'' (RIB) method. 
In the simplest rendition of this approach, we simply count how many times
a given feature of the solution appears in the different replicas.
If a structure of the solution (e.g. in the travelling salesman problem
that we will discuss in this work, a tour sequentially passing through the same
three cities) is common to all or many solvers then one may anticipate this 
structure appears in the optimal global solution. That is, augmenting the GDC
discussed above, the replicas interact effectively with one another via their correlations.
By sequentially finding common features in independent copies or replicas of the problem
and assuming these to be correct and left untouched, the system to be examined is sequentially made
smaller and easier to solve anew.
Both of the approaches that we will employ in this work may be viewed as emulating the minimization of an effective multi-replica ``free-energy''.
Schematically, as in Ref.\ \cite{ref:peter2009}, we may consider an effective free-energy 
given by

\begin{eqnarray}
\label{free-energy}
F=\sum_{i=1}^ R E_i[\phi _i]-TS[\{\phi_i\}_{i=1}^{R}]
\end{eqnarray}
where $\phi_{i}$ are the collection of coordinates that describe solver (replica) $i$, the quantities $\{E_{i}\}$ are the energies of contending solutions in the disparate replicas,
 $S$ is an inter-replica correlation functional, and $T>0$ sets a relative weight
between the sum of the intra-replica contributions ($\{E_{i}\}$) and the 
inter-replica correlations (that may, e.g., imitate the GDC, RIB or other couplings).
The detailed iterative minimization procedures that we describe in this work 
for the Traveling Salesman Problem are particular simple examples of the more general idea embodied by the minimization of the replica ensemble functional of Eq. (\ref{free-energy}). The springs in Fig.\ \ref{fig:figzero} symbolize 
inter-replica effects. For finite $T$, both intra-replica and inter-replica effects must be assuaged 
when minimizing $F$.

\section{Traveling salesman problem} \label{sec:TSP}

The Traveling Salesman Problem (TSP) is one of  the most famous problems in the entire field
of combinatorial optimization. It is often used as a benchmark for testing new optimization
approaches.
The TSP is an NP-hard problem, and as such, no algorithm has been discovered
which can solve it in polynomial time.  The problem is defined as follows:

\emph{Given a set of N cities, find the shortest tour which visits each city exactly
once and returns to the starting city.}

Because the best exact methods \cite{ref:padberg1991,ref:grotshel1991} for solving the
TSP take an amount of time which increases faster than a polynomial function of $N$,
numerous heuristic algorithms have been proposed which run much faster, but they fail 
to guarantee optimality primarily because of the drawbacks mentioned above.

There are roughly three classes of algorithms.
The first class is the greedy heuristic which gradually forms a tour by adding a new city at each step, 
such as the {\it Nearest-Neighbor algorithm} \cite{ref:johnson1997}.
In our method, we use the Nearest-Neighbor algorithm to initialize a candidate tour construction.
Briefly, the Nearest-Neighbor algorithm is given by three steps:
($1$) Select a random city. 
($2$) Find the nearest unvisited city and go there. 
($3$) Check if any unvisited cities are left? If yes, go to step $2$. If no, return to the first city.

The second class of heuristic TSP algorithms is a tour improvement approach.
A typical example is given by the {\it $k$-opt algorithm} which seeks to iteratively improve the current state 
of a tour by removing $k$ edges and replacing them in the most optimal way by random searching 
on a limited number of cities at a time.
A famous local search algorithm by Lin-Kernighan (LK) \cite{ref:lin1973} belongs to this class.
The LK algorithm is a variable $k$-opt algorithm which
decides which $k$ is the most suitable at each iteration step.
This makes the algorithm quite complex, and few have been able improve it.
A more in-depth study of the LK algorithm with possible improvements was made by Helsgaun 
(LKH) \cite{ref:Helsgaun2000}.

The third class of heuristic methods is a composite algorithm that combines the features
of the former two. A good example can be found in Dorigo and Gambardella \cite{ref:dorigo1997a}
where the authors combined the ant colony system (ACS) \cite{ref:dorigo1997b}
with the $3$-opt method to achieve strong results.
Here the ACS acts like a more sophisticated tour construction algorithm which allows 
communication (pheromones left by individual ants) between different ant solvers.
$3$-opt is the local optimizer which helps to optimize the results obtained with ACS.

\section{Geometrical distance coupling algorithm} \label{sec:MCA}

\begin{figure}[h]
\centering
\includegraphics[width=1.05\columnwidth]{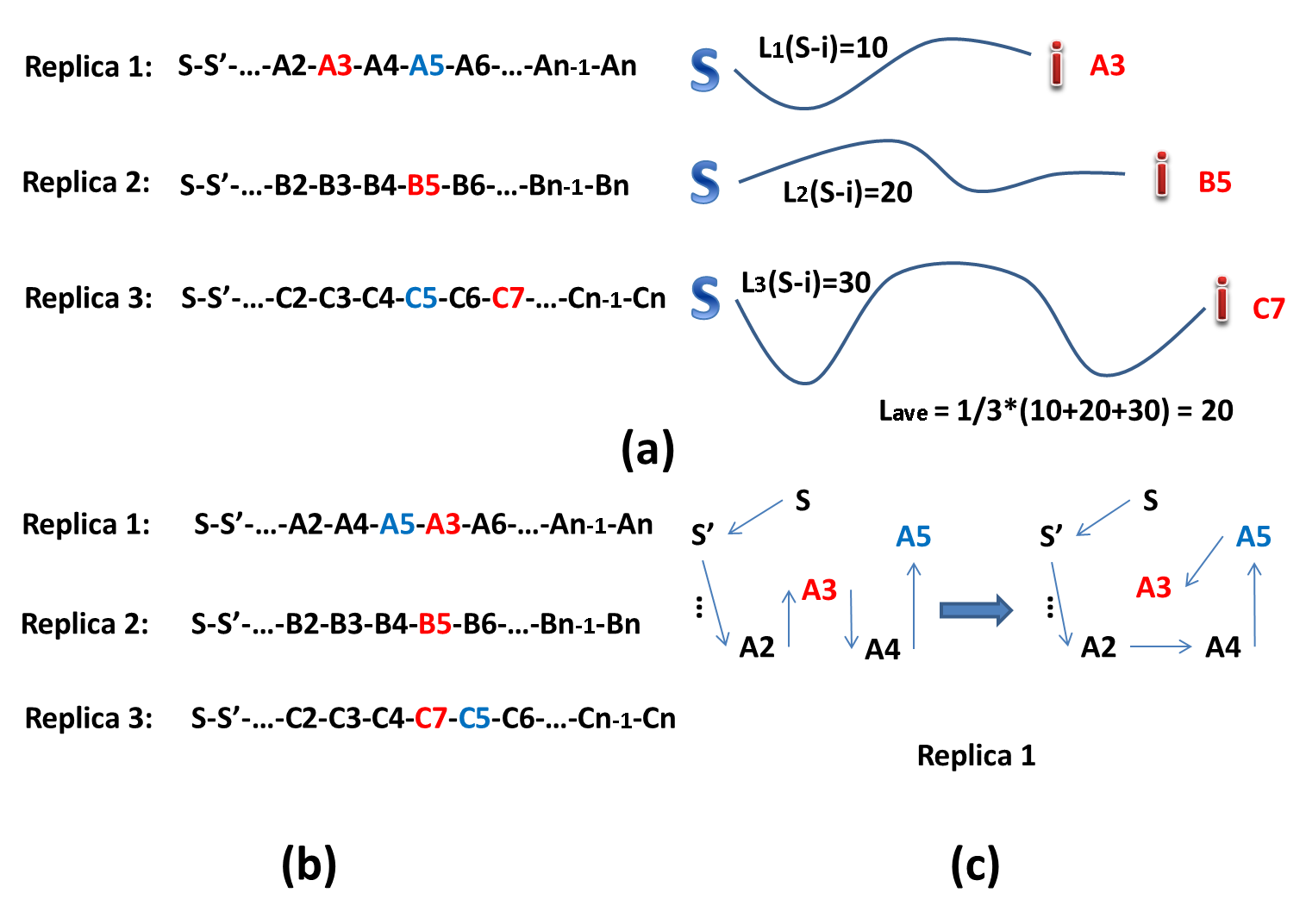}
\caption{(Color online.)
An illustration of the geometric coupling between replicas. The outcome of this basic coupling is 
that a given city (node) $i$ is moved to a position averaged over all replicas. In this example, 
there are three replicas. The link $SS'$ is common to all replicas. $S$ is a ``standard city'' that is used to calibrate distances (see Appendix).
This city is chosen at the beginning (by symmetry the choice of this city is immaterial).
(a) Three relevant replicas sharing a common edge S-S'. $i$ is the randomly chosen city. The designations A3, B5, and C7 represent the same city $i$ as it appears in replica 1, replica 2 and replica 3 respectively. In these three individual replicas, the tour length between $S$ and $i$ are 10, 20, and 30 respectively. After averaging, the city $i$ will be moved to cities with a distance of 20 away from $S$ in all three replicas. A5, B5 and C5 are the target city where A3, B5 and C7 will be inserted (see Appendix).
(b) Updated replica configurations after inserting city $i$ into the target location.
(c) A graphic depiction of the change between the initial (a) and final (b) replicas before and after the move
of city $i$ in replica 1.   
}
\label{fig:figoneee}
\end{figure}

As noted above, we use the geometric distance coupling between different solvers (or replicas) to enhance the solutions found by individual solvers. 
To demonstrate the strength of our approach and the degree to which
coupling between replicas can dramatically improve the results,
we apply our method on the bare k-opt algorithm. On their own,
sans the use of replicas, the k=2 or 3 opt-algorithms have a 
very poor performance; this makes the improvement 
using our replica based approach very clear.
Our GDC replica based approach may, in principle, 
be applied to any algorithm (not solely
k-opt). The GDC algorithm is implemented as follows:

\begin{enumerate}
\item Use the Nearest-Neighbor Algorithm \cite{ref:johnson1997} to seed all the replicas, beginning at random cities 
in different replicas to ensure some variation in the initial states.

\item Perform a variable initial number of $k$-opt steps independently on all replicas.

\item Apply the GDC step (after a given number of iterations) as follows:
\begin{enumerate}
	\item Determine the most common edge among all replicas.

	\item For replicas that share the identified common edge (see Fig.\ \ref{fig:figoneee} as well as Appendix for further discussion), 
	attempt to move a random city to its average position relative to common edge in all 
	relevant replicas. Allow the move only if the total tour length is decreased or if it is increased by less 
	than a specified tolerance.
\end{enumerate}

\item Perform a variable number of $k$-opt steps independently on all replicas.

\item Go to step $3$ until a global number of iterations is reached.
\end{enumerate}

To clarify, we start the Nearest Neighbor algorithm in different cities for each replica 
and perform an initial number of $k$-opt steps in order to guarantee that our replicas 
are not starting too close to each other in parameter space.
The GDC attempts to ensure that all of the replicas eventually converge on the same location 
in parameter space. If the replicas become clustered too closely together (as in, e.g., the cartoon of Fig. 1), they may not efficiently explore the landscape of possible solutions and fail to find the global minimum.

During the GDC step, we randomly select a sequence of cities.
We then calculate the average position of those cities relative to the most common edge 
for the replicas which contain the identified common edge.
Then, for each relevant replica, that city is plucked from its current location and placed 
in the average position.
The locally broken tour sequence is reattached in the manner described in Appendix.
If the tour length is either decreased or increased by less than the tolerance, then the move 
is accepted, otherwise it is rejected.
In this manner, the algorithm is able to locally decrease the quantity
\begin{eqnarray}
Q_{j}  \equiv \frac{1}{r} \sum_{i=1}^{r} |S_{i_{j}} - \overline{S}_{j}|.
\end{eqnarray}
In the above, $r$ is the number of the relevant replicas which share the common edge in step 3 of GDC algorithm, $1 \le j \le N$ is the city index, $S_{i_{j}}$ is the distance between city $j$ and a common edge in replica $i$, and $\overline{S}_{j}$ is the average 
of $S_{i_{j}}$ (as averaged over all the relevant $r$ replicas). By lowering $Q_{j}$ for different cities $j$, the replicas generally become more similar to one another.

\section{Results from geometrical distance coupling algorithm} \label{sec:results}

We tested our algorithm on several instances from TSPLIB \cite{ref:tsplib}
using $k = 2$ and $3$ for the $k$-opt step.
We demonstrated that within a certain range of $N$ for which $2$-opt and $3$-opt
alone almost always failed to find the global minimum, our enhanced algorithm was 
able to find it in a reasonable amount of time.
For some larger values of $N$, our algorithm also failed to find the minimum,
but it did significantly improve the $k$-opt estimate, and we believe that it could
find the minimum if mated with an appropriate optimizer which is more sophisticated 
than the standard $2$- and $3$-opt methods.

The results are summarized in Table \ref{tab:problems} where length and time values 
are averaged over $10$ runs.
The unit of the CPU time here is second.
The GDC enhanced algorithm is labeled ``2-opt GDC'' and ``3-opt GDC'' respectively, 
for the base $2$- or $3$-opt local search routine.
The parameters used here were $R=20$ replicas, $1$ geometrical distance coupling step
consisting of $N-2$ moves alternated with $1000$ to $15~000$  $k$-opt steps,
and an allowable increase in the tour length of $0.2\%$ - $1\%$ during each attempted
GDC move.
For all instances studied in Table \ref{tab:problems}, we performed $1~000~000$ initial 
$k$-opt steps independently on all replicas for step $2$ in Sec.\ \ref{sec:MCA}.
For the instances with a small number of cities (berlin52, eil51, pr76, eil76) we used
approximately $1000$ $k$-opt steps in step $4$.
For the instances with a relatively large number of cities (ch130, ts225, a280, lin 318, and att532), we invoked 15 000 k-opt moves in step 4 of the algorithm.

We allowed larger tour length increase tolerance in step $3$(b) for larger $N$ problems.
The GDC method using the $3$-opt optimization can correctly solve all examined TSPLIB \cite{ref:tsplib} problems up to $280$ cities.
If $2$-opt is applied instead of $3$-opt the maximal solvable size is $225$.  We note that neither the bare $2$- nor $3$-opt 
by themselves are not able to find the optimal solution for even the smallest of these 
examples given a comparable number $k$-opt optimization steps.
The time required to find the global optimum with the GDC step is large 
compared to the computing time for the $k$-opt.
Part of the reason is that the $k$-opt optimization is easily trapped in local minima,
but the GDC step is capable of pulling the optimizer from the local minima and having 
them explore a much broader region of the solution space.

\begin{figure}
\centering
\includegraphics[width=\columnwidth]{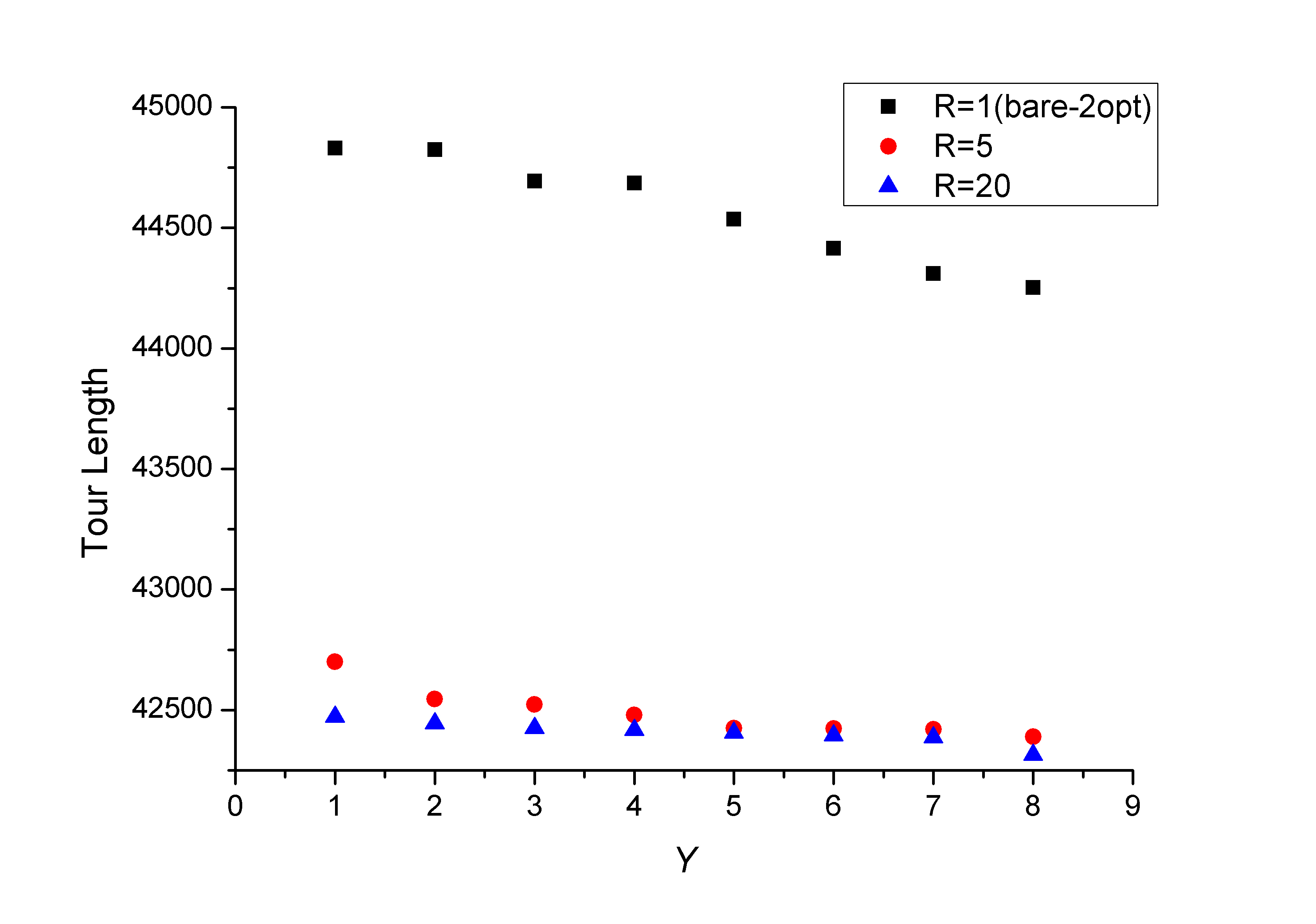}
\caption{(Color online.) The improvement of the bare 2-opt method by the use of replica coupling for the lin318 problem. The figure shows that tour lengths found by invoking R=1 (black squares), R=5 (red circles) and R=20 (blue triangle) replicas averaged over $Y \le 8$ solution attempts. 
The horizontal axis shows the results obtained by including $Y$ attempts.
Applied to the 2-opt GDC, the use of $R=20$ replicas produced better results
than the use of $R=5$ replicas.}
\label{fig:figone}
\end{figure}

\begin{figure}
\centering
\includegraphics[width=\columnwidth]{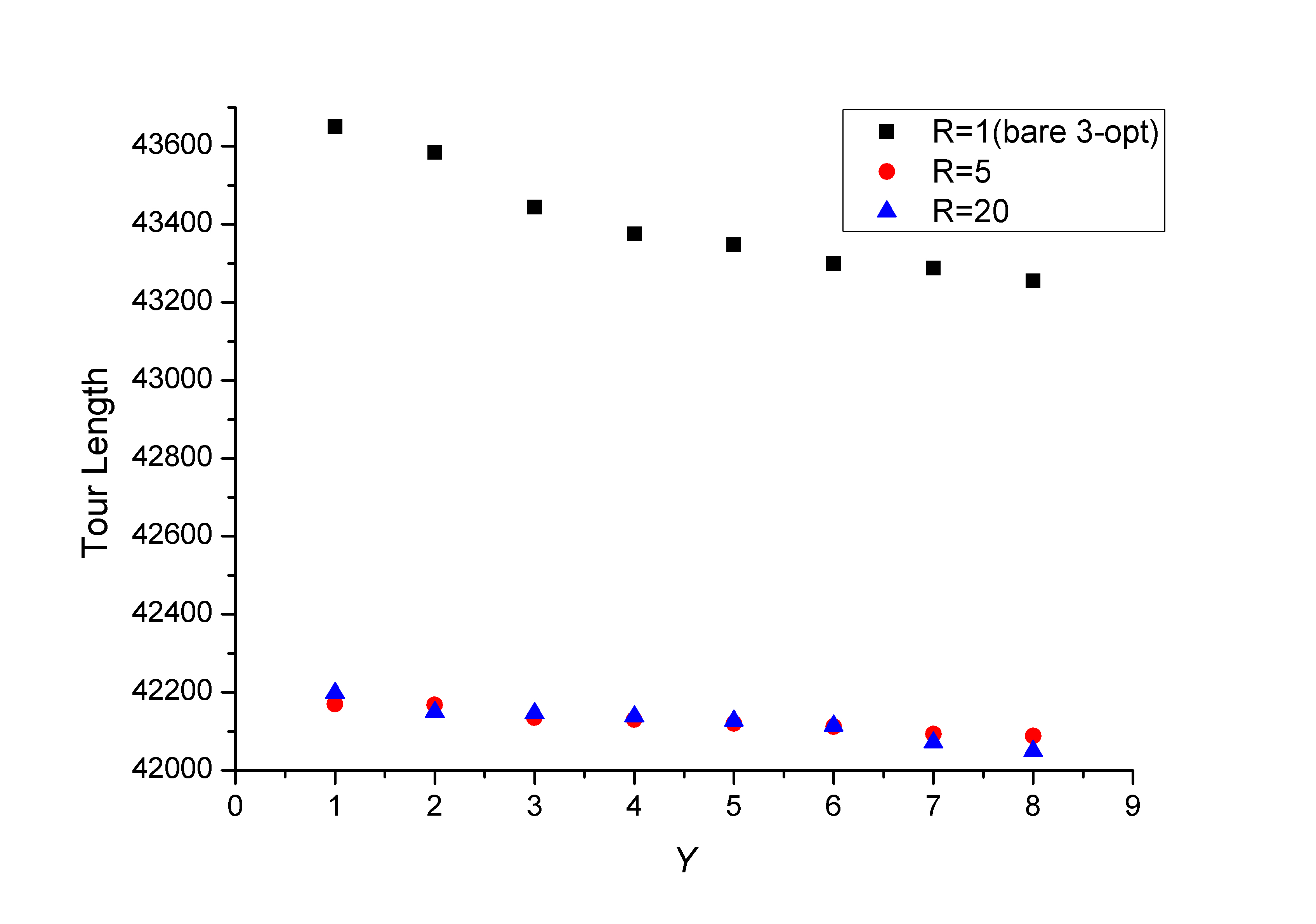}
\caption{(Color online.) The improvement of the bare 3-opt method by the use of replica coupling for the lin318 problem. 
For 3-opt GDC the average tour length for the tested range of replicas was very close, 
but when $R=20$, 
the algorithm still found a smaller tour length.}
\label{fig:figtwo}
\end{figure}

\begin{figure}[b]
\centering%
\includegraphics[width=0.95\columnwidth]{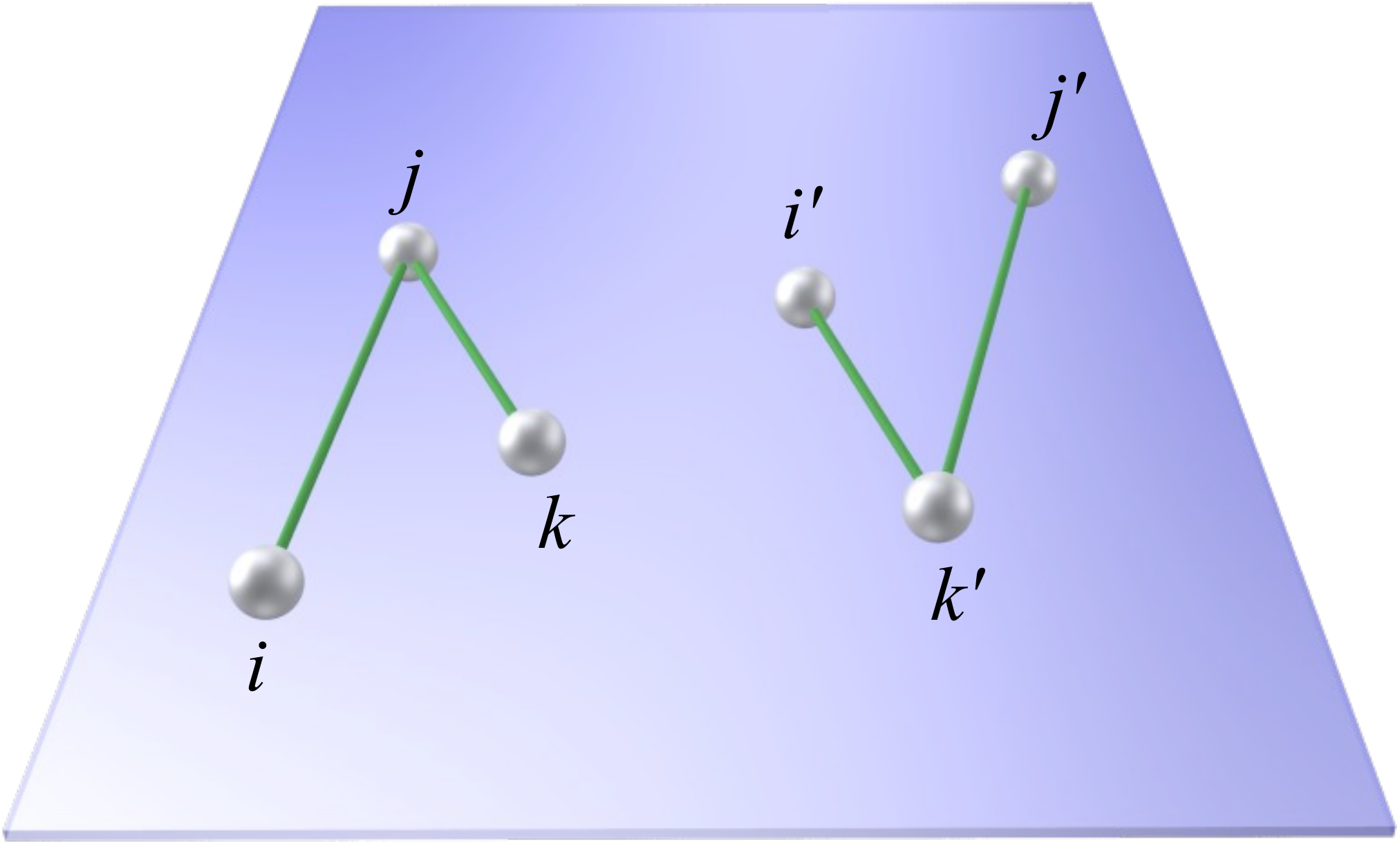}%
\caption{(Color online.)
A schematic representation common structure segments in solutions of 
the Traveling Salesman Problem.
Cities in segments $i$, $j$, and $k$ as well as $i'$, $j'$, and $k'$ are 
represented by spheres and the solved tour path follows the depicted edges connecting
the cities.}%
\label{fig:figthree}%
\end{figure}

We investigated the effect of the number of replicas used on the performance of the algorithm.
Figure \ref{fig:figone} and \ref{fig:figtwo} contrast results obtained when our GDC algorithm is applied with $R=5$ and $R=20$ replicas to improve the bare 2-opt and 3-opt respectively (we
term the resultant algorithms 2-opt GDC and 3-opt GDC) in the lin318 problem. The average tour length in the $R=5$ case using the 2-opt GDC was 42 479 whereas when using $R=20$ replicas, the 2-opt GDC yielded a path of distance 42 404. Not too surprisingly,
in both the 2-opt GDC and the 3-opt GDC, the $R=20$ replica case provides shorter tour lengths than the $R=5$ replica algorithm; this improvement with increasing $R$ is smaller for the 3-opt GDC (as the bare 3-opt algorithm is better than the 2-opt method).

\section{probabilistic replica-inference based algorithm} \label{sec:RIBA}

\begin{figure}[htb]
\centering
\includegraphics[width=1.10\columnwidth]{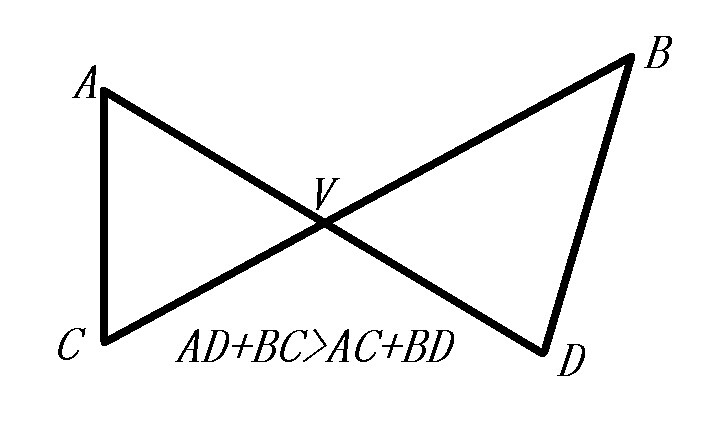}
\caption{(Color online.)
A cartoon illustrating that the minimal tour will never intersect itself. In the figure above, a tour containing the two segments $AC$ and $BD$ will always have a shorter length than a tour incorporating the same four points yet includes the segments $AD$ and $BC$ (that intersect at a point $V$). The proof of this assertion is trivial. By the triangle inequality as applied to the triangles $\Delta AVC$ and $\Delta BVD$ respectively, we have $AV + VC > AC$ and $BV + VD >BD$. Adding these two inequalities yields $AD+BC>AC+BD$. Permuting the contour segments (e.g.,
AD,BC $\to$ AC, BD) 
 to avoid crossing will always lower the total path length.}
\label{fig:figtri}
\end{figure}

Next, we show how we developed a replica-inference based (RIB) algorithm to, e.g., solve the 
lin318 test problem and the att532 test problem. The RIB algorithm is implemented as follows:

\begin{enumerate}
\item Use GDC Algorithm to seed all the replicas (the total number of replicas is R).

\item For all the replicas, calculate the probability distribution of the nodes (see equation (2)).

\item Given the probability information of the nodes, divide the tour into different parts: ``bubble'' region in which different tours appear in disparate replicas and a ``backbone'' region which is common to all replicas (see Figs. (\ref{fig:figseven}, \ref{fig:figtwelve})).

\item Keeping the common backbone region unchanged, examine different tours in the bubble regions such that when combined with the backbone path, they will lead to new viable solutions (replicas) and then pick the one (replica X) that has the shortest path. When we examine different configurations in the bubble regions we must pay attention to the ``pairing'' inside the bubbles (see Figs.(\ref{fig:figseven},\ref{fig:figtwelve}) and discussion below). We also observe that inside a given bubble there are no lines that cross as required by the triangle inequality (see Fig.\ \ref{fig:figtri}).

\item Two replica comparison (performed R times): compare the replica X found in step 4 with the R original replicas that existed prior to step 4 through steps $2$-$4$.Each time after step 4, update the replica X found in step 4. A final solution will be produced after R times of two replica comparison.

\end{enumerate}

In the up and coming, we will introduce and invoke a probability $p_{j}$ associated with each node $j$.
This probability will measure the frequency that the links to the incoming ($i$) and outgoing ($k$) associated 
with each city $j$ are the same amongst all replicas. That is,
\begin{eqnarray}
p_{j} = \frac{M_{j}}{R}.
\end{eqnarray}
$M_{j}$ is the number of times that the composite link $\langle i j k \rangle$ connecting the three cities $i$, $j$ and $k$ appears in the ensemble of $R$ 
replicas studied.

In what follows, we introduce the concept of a ``{\it bubble}'' alluded to in the algorithm above.
A ``bubble'' is, by fiat, comprised of all nodes $j$ for which the composite
links $\langle i, j, k \rangle$ are not the same across all replicas (i.e., nodes for which $p_{j}<1$). The set of such nodes must generally terminate somewhere and is linked to a backbone of nodes that have the same links in all replicas. The termination points marks the boundaries of the ``bubbles''. In the more detailed representations of the tour solutions in some of the figures
that follow, 
we will typically mark green all points $j$ for which the links $\langle i j k \rangle$
are identical in all replicas (i.e., the associated probability $p_{j}=1$).

In Fig. 6, we schematically depict typical ``one-
in-one-out" and ``two-in-two-out" bubbles (i.e., bubbles that
are attached to the common backbone by either two or four points).

\begin{figure}[h]
\centering
\includegraphics[width=0.95\columnwidth]{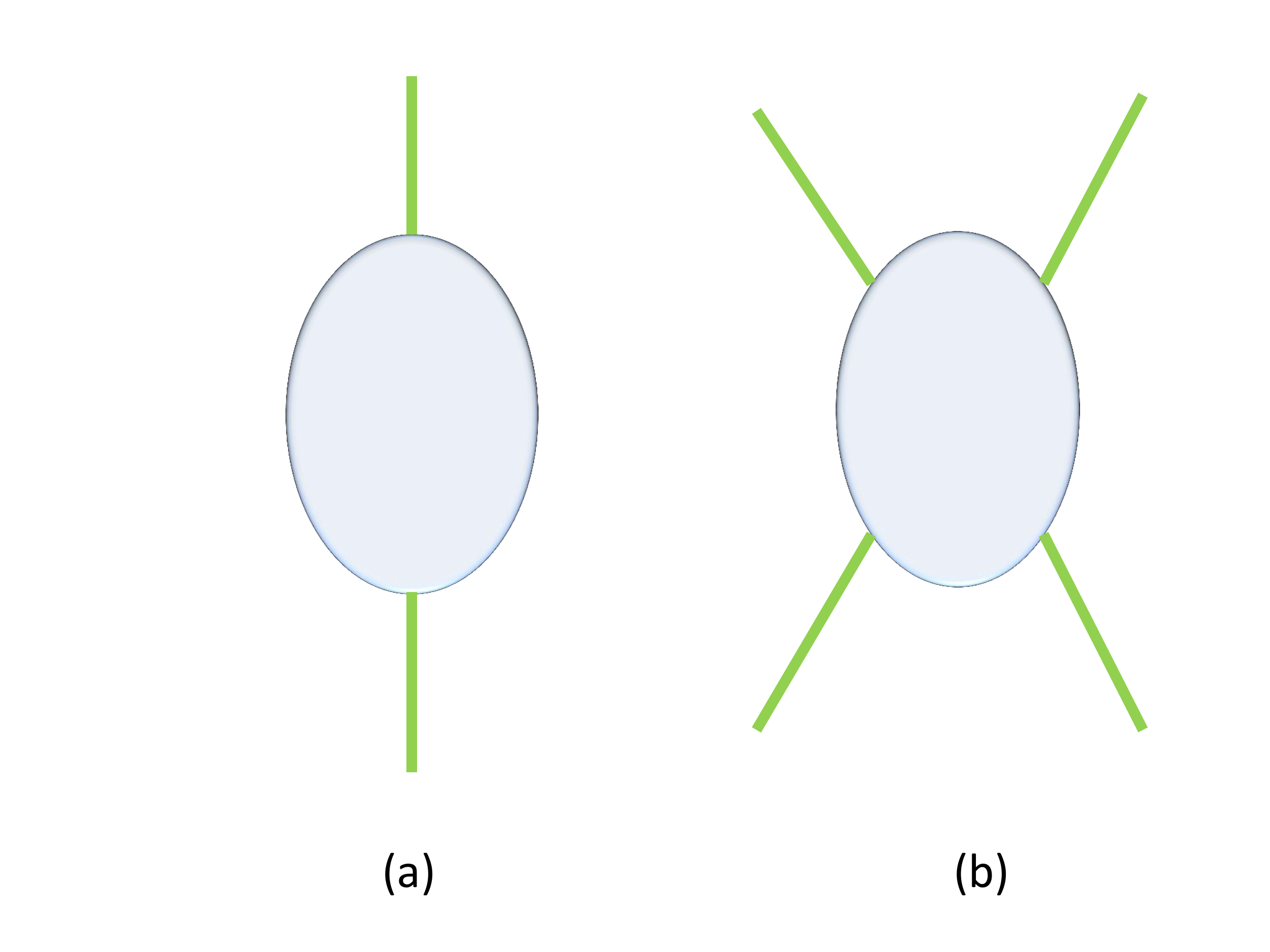}
\caption{(Color online.)
Typical ``one-in-one-out'' and ``two-in-two-out'' bubble.}
\label{fig:figonee}
\end{figure}

\section{Results of the probabilistic replica inference approach} \label{sec:results1}

We next apply our replica-inference based algorithm to solve the 
lin318 test problem from TSPLIB (see 3-opt GDC results in Table \ref{tab:problems}).
Typically, we used $R=24$ replicas.
The known true (i.e., minimal distance) tour solution is a path of length 42 029. 
Each of the $R=24$ replicas employed provided a contending solution;
the paths in each of the replicas that varied in length from 42 050 to 42 199.

\begin{figure}[b]
\centering
\includegraphics[width=0.95\columnwidth]{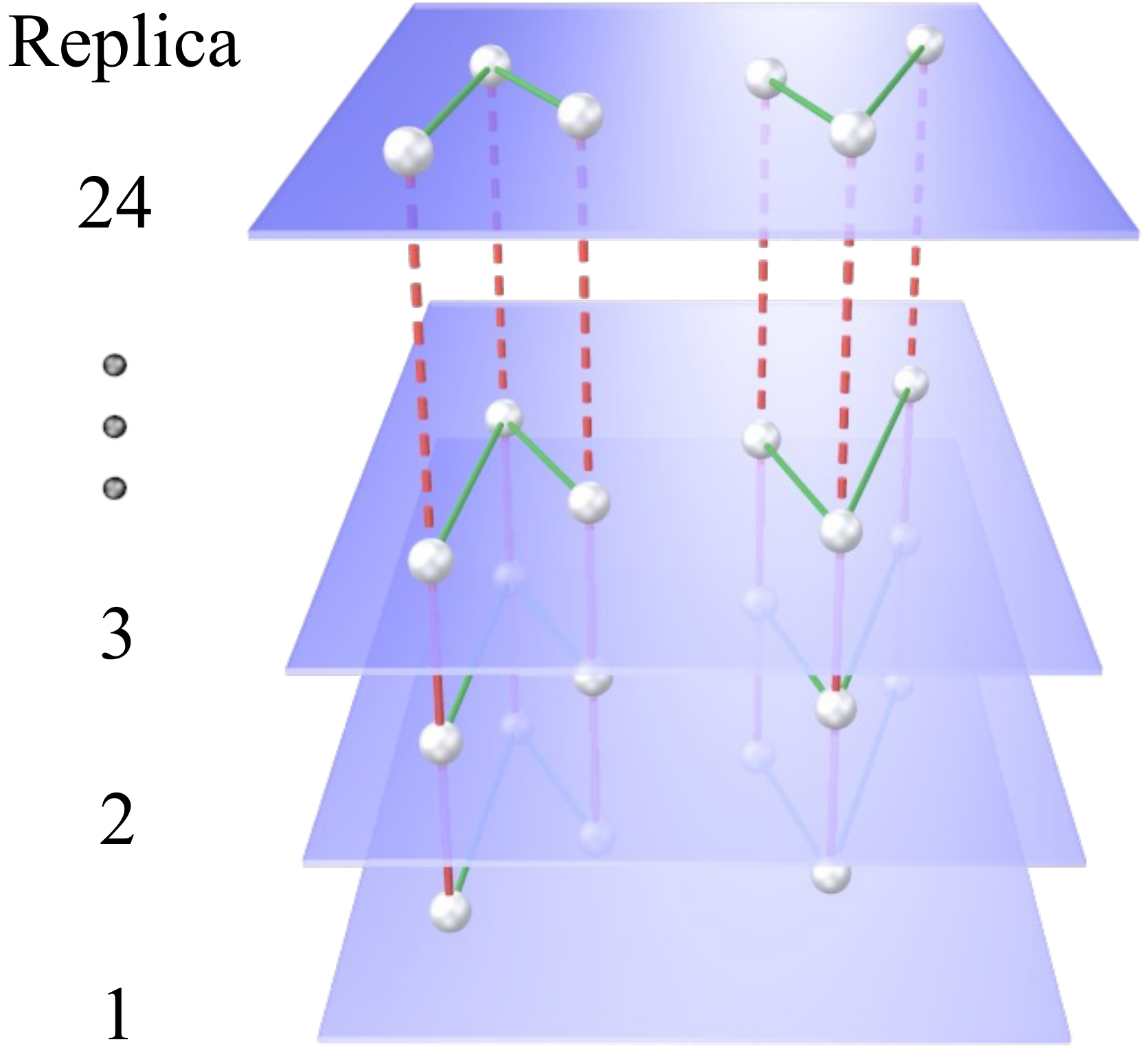}
\caption{(Color online.)
Schematic representation of common structures that appear in various replicas.
In this example, the candidate common structure contains of three nodes with two links 
between them.
When the common structure appears in \emph{all} replicas exactly, we define the probability 
to be $p_{j}=1$ ($p_{j}=24/24=1$ here).
When the structure does not appear the same in all replicas, $p_{j}<1$.
For example, if $p_{j}=1/3$, the number of replicas that contain the common structure 
is eight ($p_{j}=8/24=1/3$).}
\label{fig:figfour}
\end{figure}

\begin{figure}[b]
\centering
\includegraphics[width=0.95\columnwidth]{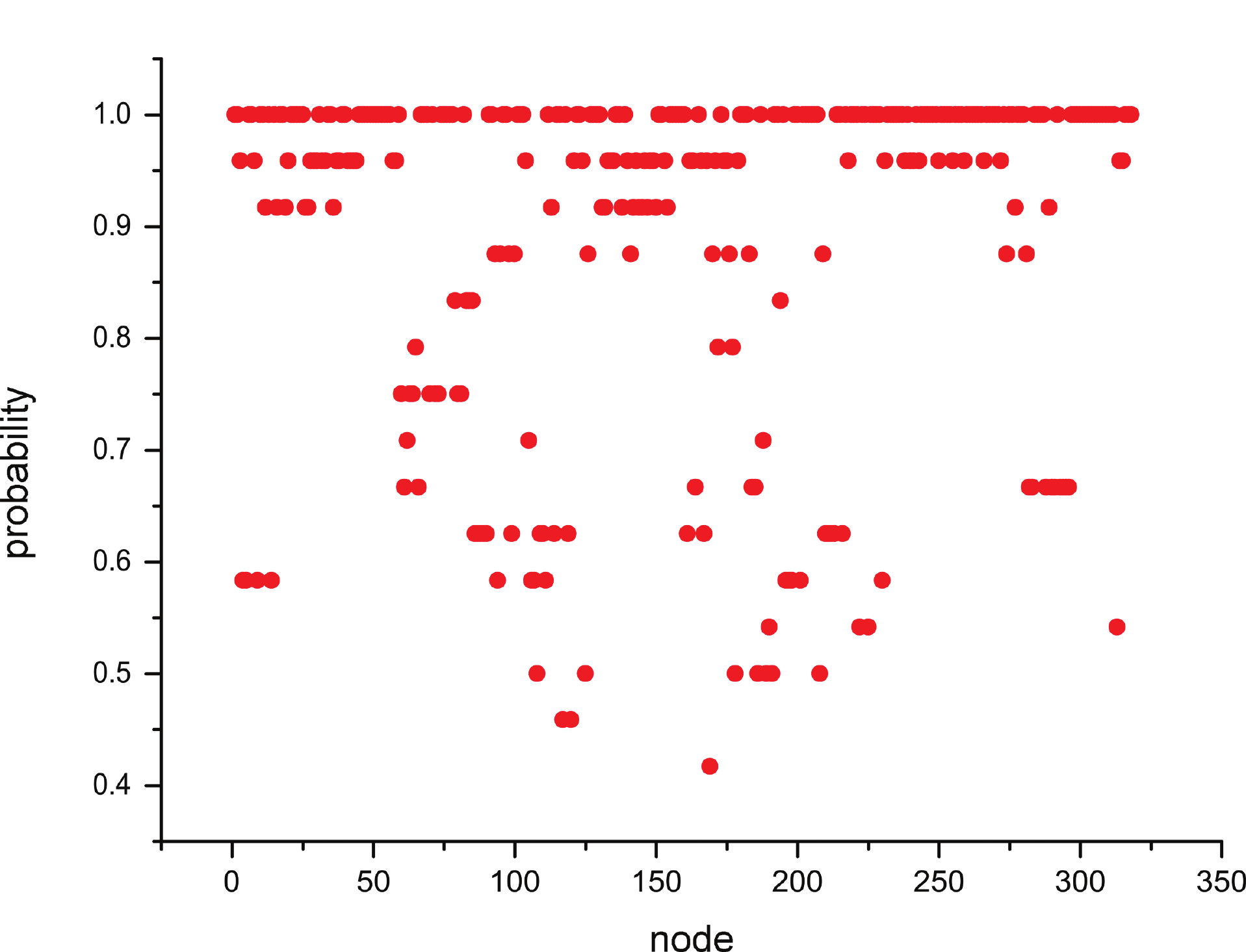}
\caption{(Color online.)
Building on the abstract representation in Fig.\ \ref{fig:figfour},
we plot the exact probability distribution $p_{j}$ (frequency of common structures identified 
in each of the different replicas) for each node in the lin318 problem from TSPLIB.
Here, there are $N=318$ nodes and $R=24$ replicas.
Every node has two adjacent neighbors, so we define $p_{j}$ as the number of times a common 
structure occurs (\ie{}, the same pair of neighbor cities are connected to node $j$) 
among the replicas divided by the total number of replicas.}
\label{fig:figfive}
\end{figure}

A significant fraction of the $R=24$ configurations produced by the GDC algorithm
(discussed in sections \ref{sec:MCA}, \ref{sec:results}) when it is applied for each of the $R$ replicas 
share three-site configurations of the type shown in Fig.\ \ref{fig:figthree}. Schematically, the composite link $\langle i j k \rangle$ may be shared amongst numerous replicas
as is illustrated in 
Fig.\ \ref{fig:figfour}.
To quantitively measure this similarity, we employ (as we have alluded to in sections \ref{sec:MCA}) a probability distribution $p_{j}$ based on the these link patterns $\langle i j k \rangle$ associated with each node $j$.
That is, in every replica each node has two adjacent nodes, so 
$p_{j}$ is defined as the max number of replicas for which $j$ shares the same neighbors 
(where $j$ is in the middle) divided by total number (R) of replicas ($R=24$ in our example calculation here).
In Fig.\ \ref{fig:figfive}, we show a representative plot the probability distribution 
for different nodes on the lin318 problem.

\begin{figure}
\centering
\includegraphics[width=0.95\columnwidth]{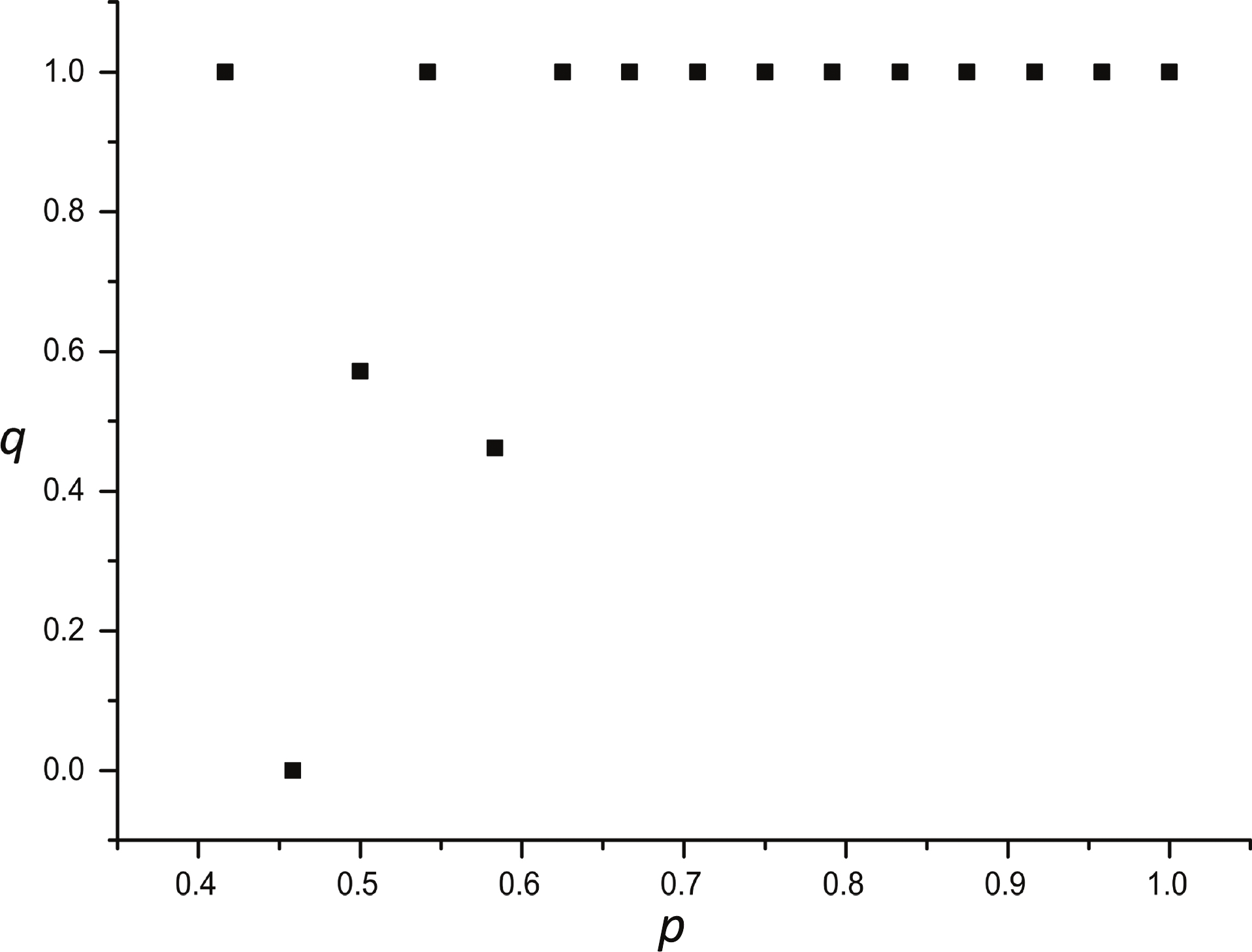}
\caption{(Color online.) 
The vertical axis is the fraction ($q$) of two links from neighboring sites that impinge on a given node ($j$) found by the replicas that appear in the optimal (i.e., shortest tour) length solution. The horizontal axis is the probability ($p_{j}$) of finding this common set of two neighbors connected to the given node $j$; this probability $p_{j}$ is identical to the vertical axis in Fig.\ \ref{fig:figfive}. As this figure illustrates for sufficiently large values $p_{j}$, essentially all of the links found by many replicas also appear in the true optimal solution. }
\label{fig:figsix}
\end{figure}

We wish to take advantage of the information contained in all of replica tours.
Toward that end, we observe in Fig.\ \ref{fig:figfive} that $162$ out of $318$ nodes
(approximately $50\%$ of the nodes) have a probability of $1$.
We conjectured that the common structures for nodes which have a probability equal 
to $1$ are the same as that from the known optimal solution.
To test whether it was the case, we made a plot of Fig.\ \ref{fig:figsix}.
In Fig.\ \ref{fig:figsix}, $p_{j}$ is a given fixed probability $p_{j}$
defined above when averaged over all replicas.
Then we looked at the set of all links $\langle i,j,k\rangle$ having that probability 
$p_{j}$.
$q$ is the fraction of these links that appear in the optimal solution; 
we then plot $q$ versus $p_{j}$ in Fig.\ \ref{fig:figsix}.

\begin{figure}
\centering
\includegraphics[width=0.95\columnwidth]{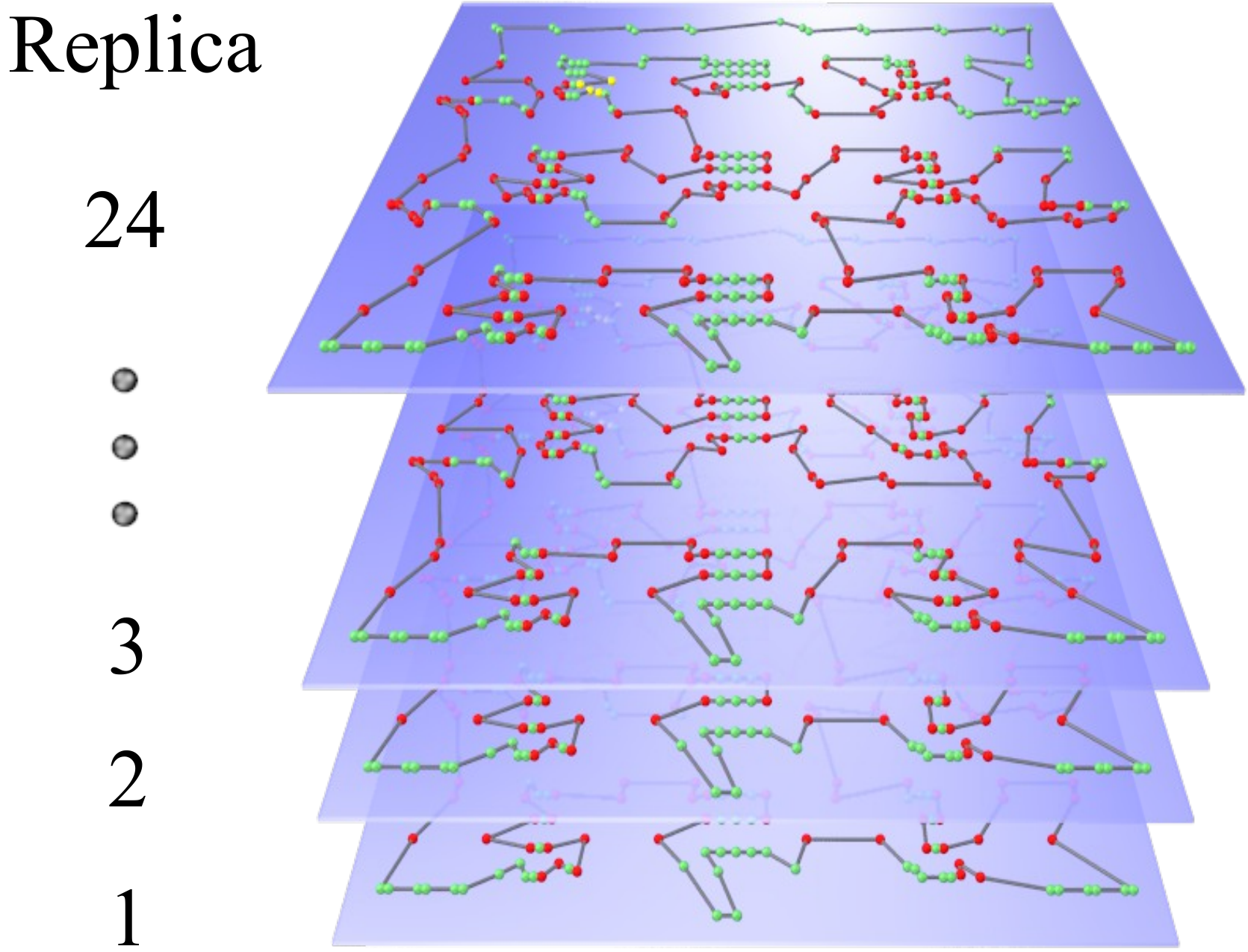}
\caption{(Color online.)
We define a ``bubble'' as a set of nodes where the neighbor cities
differ among the replicas.
Here, green nodes denote the nodes which have identical neighbors in all $R$ 
replicas ($R=24$ here); we define these nodes to have a probability $p=1$.
In this example, there are two distinct bubbles with nodes that are, respectively, depicted in this figure by two different colors- i.e., ``yellow'' and ``red'' spheres.
The configurations inside the bubbles are different for each replica while the tour
sections outside the bubbles are the same for all $24$ replicas.}
\label{fig:figsevenn}
\end{figure}

\begin{figure}
\centering
\includegraphics[width=1.05\columnwidth]{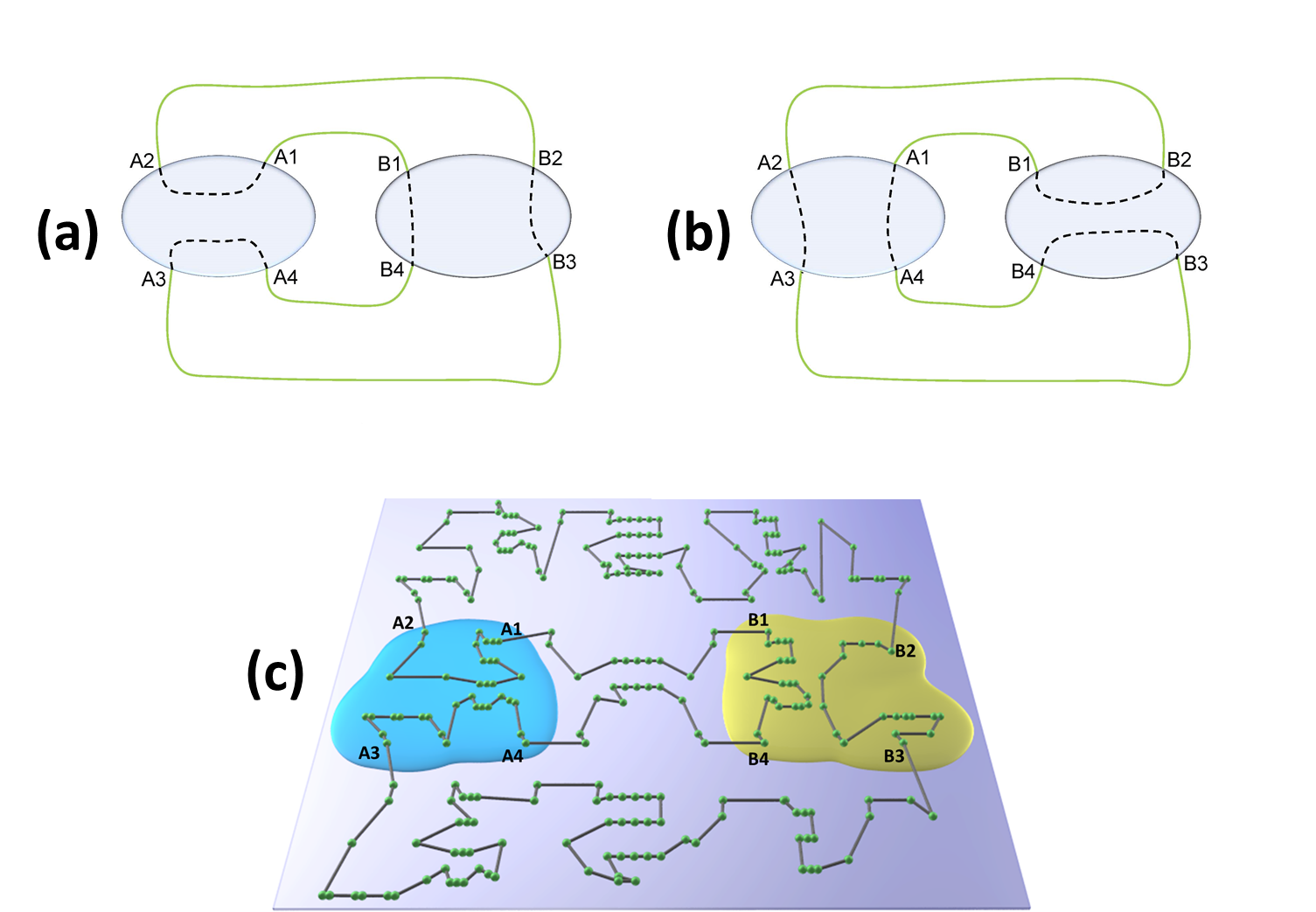}
\caption{(Color online.)
A schematic top view of Fig.\ \ref{fig:figsevenn}. (a) one possible pairing inside the bubbles (b) the other possible pairing inside the bubbles. The green solid lines outside the blobs refer to the common structures shared by all of the 24 replicas while the dotted line inside the blobs denote the various possible bubble tours. The nodes (A1, A2, A3, A4, B1, B2, B3, and B4) are located on the boundaries of the shown bubbles. (c) a concrete example of (a). }
\label{fig:figseven}
\end{figure}

Perusing Fig. 9 (associated with the lin318 problem), we observe 
that links $\langle  i, j, k \rangle$ with inter-replica frequency
$p=1$ (i.e., links that consistently appeared in all replicas) 
indeed appeared in the optimal tour. Furthermore, numerous links 
$\langle i, j , k \rangle$ with $p<1$ (i.e., those which were not consistent 
across all replicas) also appeared in the optimal shortest tour solution. 
In what follows, we introduce the concept of a ``bubble'' as it pertains to
the current problem. A ``bubble'' is, by fiat, comprised of all nodes $j$ for which the
links $\langle i, j, k \rangle$ are not the same across all replicas (i.e., nodes for which $p<1$). The set of such nodes must generally terminate somewhere and is linked to a backbone of nodes that have the same links in all replicas. The termination points marks the boundaries of the ``bubbles''.

\begin{figure}[h]
\centering
\includegraphics[width=1.05\columnwidth]{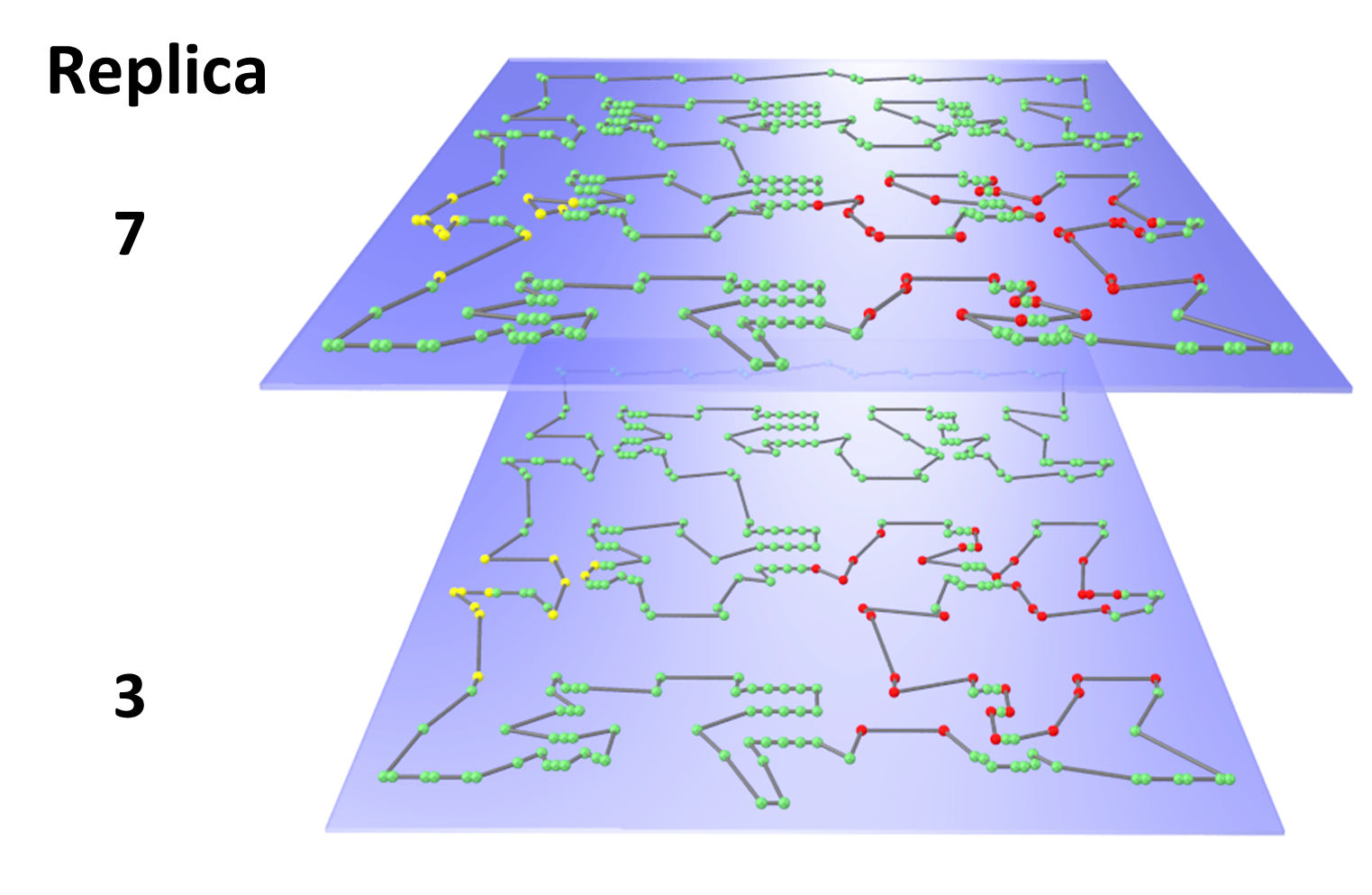}
\caption{(Color online.)
A specific illustration of how our method is applied to two particular replicas in our GDC algorithm in Sec.\ \ref{sec:MCA}.
The top place denotes the outcome of replica number 7 in our simulations while the bottom plane shows replica number 3. 
Green nodes are those nodes that have identical links in the set of all replicas (as in Fig.\ \ref{fig:figfour}, \ \ref{fig:figfive}). That is, nodes that are colored green have identical neighbors in all replicas. The remaining nodes with links that differ between the disparate replicas form separate ``bubbles'' attached to the backbone of common (green) nodes. We mark the nodes in the different ``bubbles'' by different colors. The yellow and red nodes form two bubbles where the tour configurations in the individual replicas differ. Amongst the two replicas shown, the shorter path in the bubble formed by the yellow nodes appears in replica 3. This intra-bubble configuration may be implemented in replica 7 to replace the original one shown. Once this transfer is done, the optimal tour (shown in Fig.\ \ref{fig:fignine}) results.}
\label{fig:figeightt}
\end{figure}

In the replica-based inference approach, the common structures 
with $p=1$ are left untouched.
We aim to solve the smaller and less difficult bubble problems separately instead 
of the entire tour map.
In Figs. \ref{fig:figsevenn} and \ref{fig:figeightt}, nodes (represented as spheres in a
connected tour map) whose probability is $p=1$ are colored green and those with $p<1$ 
are colored red or yellow.

We now turn to step 3 of our algorithm. By definition, the bubbles encompass the same set of nodes in all replicas.
That is, if there is at least one replica in which a red (or yellow) node $a$ is attached to another 
red (or yellow) node $b$, then $a$ and $b$ will lie in the same bubble for all replicas.
In the lin318 problem, we obtained two bubbles by comparing the $24$ 
replicas to each other. This is illustrated in Fig.\ \ref{fig:figsevenn}. (Different colors refer to different bubbles.)

Next, we apply step 4. As mentioned previously, the green nodes in Fig.\ \ref{fig:figsevenn} remain unchanged.
We then solved for  the optimal tour  inside the two identified bubbles for this problem.
The shortest intra-bubble tour for the larger bubble (red nodes) is $26~591$ (from replica $7$).
The shortest intra-bubble tour for the smaller bubble (yellow nodes) is $578$ (this also appeared in replica $7$).
Although we cannot find the optimal solution for lin318 problem at this stage, the current best tour
length is $42~050$.

\begin{table*}
\begin{tabular}{ |p{1cm}|p{1cm}|p{1cm}|p{1cm}|p{1cm}|p{1cm}|p{1cm}|p{1.2cm}|p{1.2cm}|p{1.2cm}|p{1.2cm}|  }
\hline
\multicolumn{3}{|c|}{Benchmark problems}&\multicolumn{4}{|c|}{Bare $k-opt$ algorithms unable to find optimal path}&\multicolumn{4}{|c|}{Results from our new method that couples these $k-opt$ algorithms} \\
\hline
& & & $2$-opt &  & $3$-opt &  & 2-opt GDC &  & 3-opt GDC & \\
\hline
Problem & Cities & Optimal length & Length & CPU time & Length & CPU time & Length & CPU time & Length & CPU time \\
\hline
berlin52 & 52 & 7542 & 7721 & 0.035 & 7606 & 4.3 & 7542 & 2.31 & 7542 & 1.62 \\
eil51 & 51 & 426 & 433 & 0.023 & 429 & 13.41 & 426 & 6.82 & 426 & 39.36 \\
pr76 & 76 & 108159 & 110875 & 0.057 & 109096 & 27.32 & 108280 & 8.511 & 108159 & 399.57 \\
eil76 & 76 & 538 & 553 & 0.05 & 544 & 29.35 & 538 & 50.3 & 538 & 184.63 \\
ch130 & 130 & 6110 & 6354 & 5.09 & 6232 & 8.377 & 6110 & 1003 & 6110 & 410 \\
ts225 & 225 & 126643 & 128103 & 7.08 & 126885 & 110.4 & 126643 & 3398.6 & 126643 & 76.44 \\
a280 & 280 & 2579 & 2701 & 8.88 & 2642 & 143.43 & 2615 & 1500 & 2579 & 7807.8 \\
lin318 & 318 & 42029 & 44473 & 7.66 & 43347 & 48.57 & 42423 & 15120 & 42124 & 28080 \\
att532 & 532 & 27686 & 29338 & 10.88 & 28447 & 60.00 & 28607 & 20220 & 28035 & 40770 \\
\hline
\end{tabular}
\caption{TSP performance and results using $2$-opt, $3$-opt, 2-opt GDC, and 3-opt GDC 
on a selection of TSPLIB instances. k-opt GDC are results from the current work.
The values of tour lengths and CPU times are averaged over $10$ runs.
For the studied instances, $2$-opt and $3$-opt always failed to find the global minimum alone.
With geometrical distance coupling (see Sec.\ \ref{sec:MCA}), our 2-opt GDC 
algorithm found the global minimum up to $N=225$ cities, and 3-opt GDC found the optimal tour 
up to $N=280$.
Although neither 2-opt GDC nor 3-opt GDC found the optimal solution for lin318,
they significantly improved the base $2$- or $3$-opt estimate.
The percentages above the optimum for lin318 was $5.8\%$ for $2$-opt and $3.1\%$ for $3$-opt
which was reduced to $0.94\%$ and $0.22\%$ for 2-opt GDC and 3-opt GDC, respectively.}
\label{tab:problems}
\end{table*}

Caution must be exercised in choosing the optimal ``intra-bubble'' tour among all the relevant replicas.
We mandate that the resultant tour is a valid TSP solution (i.e., we need to make certain that (i) the tour visits each node exactly once inside any bubble
and that (ii) the formed global tour forms a closed path).
To that end, we should consider the {\it Pairing} between two nodes alludes here to 
the circumstance that these two nodes are continuously connected 
to each other by an intra-bubble segment (see Figs. (\ref{fig:figseven}, \ref{fig:figtwelve})).
Figs. (\ref{fig:figseven}, \ref{fig:figtwelve}) constitute top views of the tours depicted in Figs. (\ref{fig:figsevenn}, \ref{fig:figten}) where the common backbones and regions with differing node paths (``bubble'') are made clear.
The blobs in Fig.\ \ref{fig:figseven} schematically denote bubbles. The green solid lines (backbones) are the tour path formed only by common structures (which we do not want to change). The dotted lines inside the blobs are possible tour paths (we want to find the shortest such paths). Within the blobs there might be some common structures between the different replicas.
Among all of the 24 replicas there were the two possible ways of pairing for the bubble nodes at the boundary.
The bubbles in Fig.\ \ref{fig:figseven} are of ``two-in-two-out'' type.
That is, the full tour will enter and exit each bubble twice. 
As discussed above, when we pick the optimal solution for each bubble and combine them together to form a new global solution we must make it certain that the tour is still valid.

\begin{figure}
\centering
\includegraphics[width=\columnwidth]{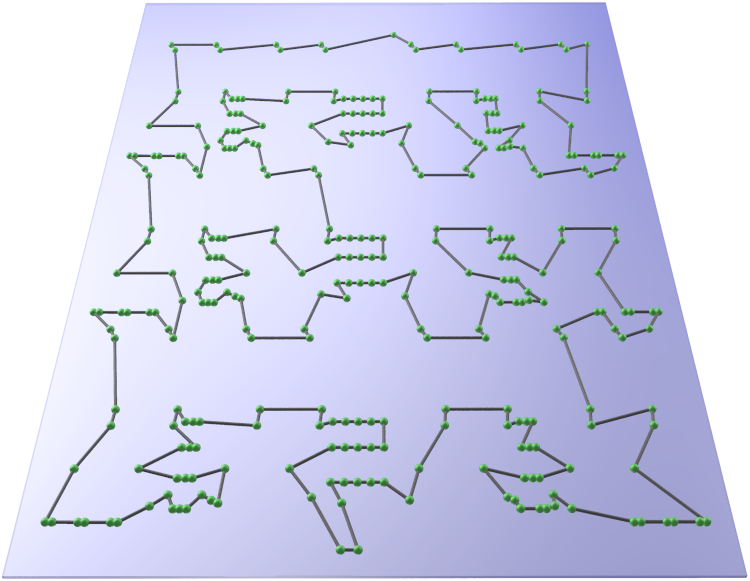}
\caption{(Color online.)
Optimal solution for lin318 from TSPLIB as discussed in Fig.\ \ref{fig:figeightt}.
Following this transfer of the tour segment inside the bubble from replica 3 in Fig.\ \ref{fig:figeightt}, the new replica 7 attains the lowest distance optimal tour for the lin318 problem.}
\label{fig:fignine}
\end{figure}

\begin{figure}
\centering
\includegraphics[width=0.95\columnwidth]{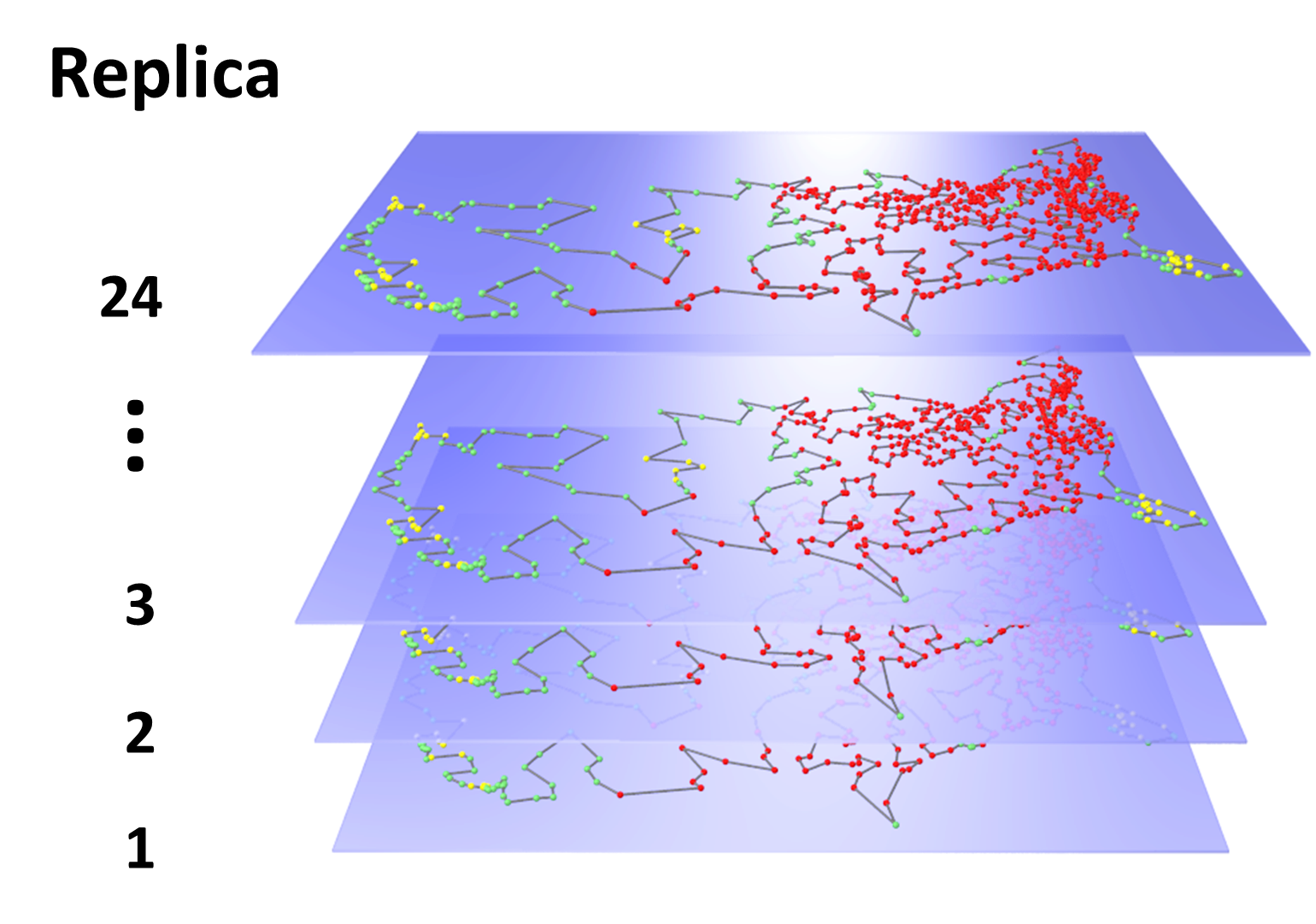}
\caption{(Color online.)
An all replica comparison that was used to produce the common (green) backbone of links for the 532 node att532 problem. In this example, we found a total of six bubbles attached to the common backbone. Five of these bubbles were of the ``one-in-one-out'' type (denoted yellow above). The nodes in the more challenging ``three-in-three-out" type bubble are marked red.
}
\label{fig:figten}
\end{figure}

\begin{figure}[h]
\centering
\includegraphics[width=0.95\columnwidth]{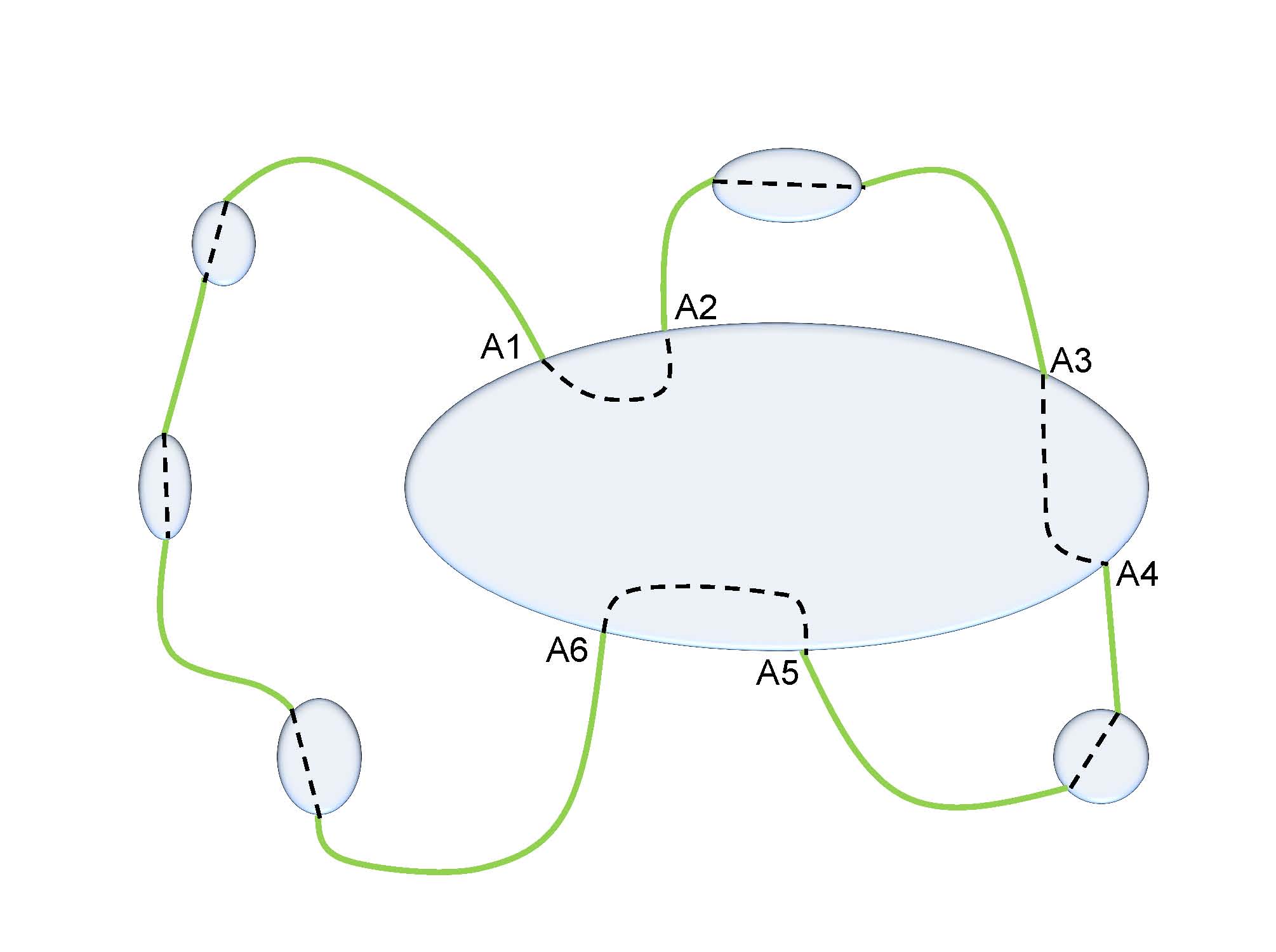}
\caption{(Color online.)
A schematic top view of Fig.\ \ref{fig:figten}. The green solid lines denote the common tour path between the replicas. The dotted line inside the blobs denote the possible various bubble tours. The nodes (A1, A2, A3, A4, A5, and A6) are situated on the periphery of the ``three-in-three-out'' bubble marked red in Fig.\ \ref{fig:figten}.  
}
\label{fig:figtwelve}
\end{figure}

Step 5:  We were able to decrease the old tour length for replica $7$ by using just two replicas.
In doing so, we find that we can decrease the tour length from $42~050$ to $42~029$
by swapping common bubble appearing in both replicas $3$ and $7$ and having the same ``in-out'' pairing.
The bubbles being swapped are of the ``two-in-two-out'' type with the same pairing (see Fig.\ \ref{fig:figseven}). So we can safely swap these two bubbles.
The results are shown in Figs.\ \ref{fig:figeightt} and\ \ref{fig:fignine}.
The total tour length associated with the common bubble in replica $3$ is $4042$ compared to $4063$ for
replica $7$ with a difference of $21$.
Upon swapping the lower distance tour in the smaller bubble from replica 3 with the existing one
in replica 7, replica $7$ attained the ideal optimal tour length of length $42~029$- the correct solution of the
lin318 problem.
The resulting tour is depicted in Fig.\ \ref{fig:fignine}.

\begin{figure}[h]
\centering
\includegraphics[width=0.95\columnwidth]{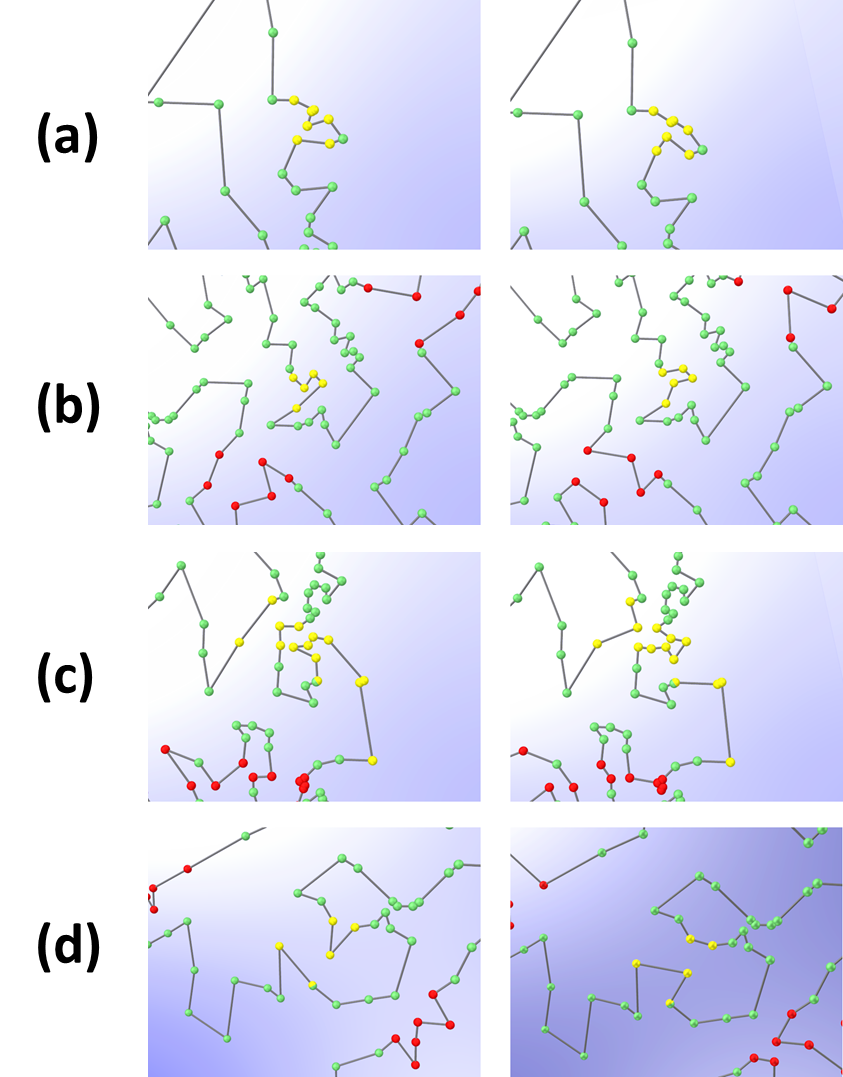}
\caption{(Color online.)
A comparison between replica 17 and other replicas in the att532 problem (see Figs.\ \ref{fig:figten} and \ref{fig:figtwelve} for notation and color convention). (a) A side by side comparison between replica 17 (left) and replica 1 (right). Once the shorter intra-bubble path (marked yellow) from replica 1 is implemented in replica 17, the total tour length in replica 17 is reduced by a distance difference of size 8.  (b) A similar comparison (for a region with another one-in-one-out ``yellow bubble'' different than that shown in panel (a)), between replica 17 (left) and replica 1 (right). Coincidentally, here also, swapping the intra-bubble tour in replica 17 with the shorter one found in replica 1 further lowers the total length by 8. (c) A further analogous comparison between replica 17 (left) with replica 10 (right). Replacing the initial other intra-bubble tour in replica 17 by the shorter one found for this bubble in replica 10 leads to a further lowering the tour length by 26. (d) A comparison between replica 17 (on left) with replica 23 (right) for a fourth ``one-in-one-out" bubble in Figs. (\ref{fig:figten}, \ref{fig:figtwelve}). Using the shorter intra-bubble tour found in replica 23 instead of that initially found in replica 17 leads to a further reduction of the tour length in replica 17 by 14.}
\label{fig:figthirteen}
\end{figure}

We also applied our replica inference method to further optimize the solutions obtained by the GDC algorithm for the att532 problem \cite{ref:tsplib}. 
In Fig.\ \ref{fig:figten}, all 24 replicas were employed to produce the common backbone (``green'') nodes and bubbles (differing tour regions attached to the common backbone) as we did for the lin318 instance. There were six bubbles in total. Five of these bubbles were of  the ``one-in-one-out'' type while the rest were of the ``three-in-three-out'' type (see Fig.\ \ref{fig:figtwelve}). In all of the replicas, the "in-out" pairing was between the very same sets of node pairs. For the five smaller bubbles,  by virtue of
their minute size, brute force minimization quickly produced to find the optimal solution. For the ``three-in-three-out'' bubble we inserted the shortest path result (that obtained from replica 17) among the 24 replicas. These steps led to a solution for the att532 problem having a total tour length of 27 937 as compared to the average replica result of 28 035.
We tried to decrease the tour length of the big bubble of ``three-in-three-out'' type further by invoking replica pair comparisons. This iteratively led to a replacement of the original three-in-three-out bubble to many far smaller bubbles. We then investigated whether we can optimize the original big bubble by swapping these smaller bubbles between replica 17 and others.  Adopting some smaller bubble solutions from replicas 1,10, and 23 respectively, this set of sequential minimization and inference operations  led to a better solution having a tour length $27~881$. The process is illustrated in Fig.\ \ref{fig:figthirteen}. The length of the resultant contending solution is 27 881. (The known optimal shortest path for att532 has a tour length equal to $27~686$.)

Although we still cannot find the global minimum for att532, the solution of 27 881 produced by GDC algorithm and bubble method togother is only $0.7\%$ above the optimum. More importantly, by employing inference, we were able to reduce the large problem involving a minimization of the path of all nodes in the graph to a set of smaller problems involving disparate ``bubbles''. Other than the ``three-in-three-out'' bubble the rest tour configuration was found to be the same as the known optimal solution.

To better understand the difference between our solution and the known optimal solution, we compared our replicas to the known optimal solution. Perusing Fig.\ \ref{fig:figtwelve},  we find that despite of the same in-out ``pairing'' occuring associated with identifying the locations where the tour went in and out of each bubble, the optimal solution tries to go from node (on boundary) A6 to A5 more directly while our replicas always intend to go north from city A6 and finally reach A5 after visiting numerous nodes.

\section{Conclusion} \label{sec:conclusion}

We introduce {\it a general method for improving known algorithms.}
(i.e., the $2$-opt and $3$-opt of the TSP \cite{ref:johnson1997}). Although, for definitiveness,
we focused on the TSP, the premise of underlying our method is very general and it may, in principle, be applied to many other problems. The core concept of our approach is that of {\it coupling between independent solvers}
(see Fig.\ \ref{fig:figzero} and Eq. (\ref{free-energy})). Such a coupling between members of an {\it ensemble} of solvers (or ``replicas'')
that collectively seek to find an optimal solution
 may substantially improve the convergence to the correct answer
as compared to the prevalent single replica algorithm.
This coupling may be introduced amongst solvers of many types (with these solvers obtained by any previously known algorithm). 
We underscore and reiterte that by couplings these solvers, we may, very significantly, improve earlier results. 
In the context of the TSP, we demonstrated that geometrical distance and probabilistic inference coupling 
between otherwise independent replica solvers allow local solvers to flee from false local minima. By doing this, we obtained 
optimal TSP tours even when single solvers were unable to find the correct answer. As an example, we showed 
that while the bare 3-opt method fails to solve for the examined TSP problems with 
more than 50 cities (see Table \ref{tab:problems}), {\it by invoking replicas, the 3-opt method can accurately solve 
problems up to size of 318 cities} (see Fig. \ref{fig:figeightt}). We furthermore obtained nearly optimal solutions even for larger 
systems. For instance, in the att532 example the replica method led to a solution with $0.7 \%$ increase in tour length relative to the minimum (see latter part discussion of Section \ref{sec:results1}). Thus, the inter-replica coupling indeed led to a substantial improvement. Unlike genetic algorithms that involve no such coupling and simply randomly swap the city nodes among different contending solutions, our algorithms makes use of replica correlations and inference. Genetic algorithms fail to solve problems as complicated as those we do. To our knowledge, the currently best genetic algorithm \cite{ref:ga2010} already falters in atempting to correctly find the minimal path for a 76 city tour (``pr76'') that we readily solved here (see Table \ref{tab:problems}) and successfully went to far larger city tours. In conclusion, we introduce a new replica based approach that may be applied to disparate problems beyond the confines of the particular TSP problem solved here and the clustering and image segmentation challenges addressed in \cite{ref:peter2009,dandan1}. Our core idea is that even algorithms that are simple may be much more potent
once {\it inter-replica interactions} and {\it inference} are invoked.


\section*{Acknowledgment}

This work was supported by NSF grants DMR-1411229 and DMR-1106293.

\appendix

\section{Geometrical distance coupling step}  \label{app:mics}

The geometrical distance coupling step is applied ``on top of''
a base TSP solver.  Generally speaking, the idea is to alternate
optimization of the local solver ($k$-opt in the current work) 
with the GDC step to induce the algorithm to escape local minima 
and enhance the chances of finding the globally optimal tour solution.
GDC seeks to utilize the distance information implicitly contained 
in multiple TSP solvers to enhance the optimization performed by a base 
algorithm.

For illustration purposes, the following discussion uses only five replicas
which are labeled $a$, $b$, $c$, $d$, and $e$.
We then represent a candidate tour solution as a string of $N$ cities,

Replica $a$: tour: $a_1, a_2, a_3, \ldots, a_N$

Replica $b$: tour: $b_1, b_2, b_3, \ldots, b_N$

Replica $c$: tour: $c_1, c_2, c_3, \ldots, c_N$

Replica $d$: tour: $d_1, d_2, d_3, \ldots, d_N$

Replica $e$: tour: $e_1, e_2, e_3, \ldots, e_N$

\noindent where each replica tour correspondes to a permutation of the integers $(1, 2, 3, 4, \ldots, N)$.
The GDC algorithm is given by the following steps:

\begin{enumerate}
\item Determine the most common edge among all replicas:

\begin{enumerate}
	\item Randomly select a ``standard'' reference city from $2, 3, 4, 5, \ldots, N-1$.
	Cycle each replica order so it includes the standard city as the first element.
	The five example replicas have the following configurations:

  Replica $a$: $S, a_2, a_3, \ldots, a_N$

  Replica $b$: $S, b_2, b_3, \ldots, b_N$

  Replica $c$: $S, c_2, c_3, \ldots, c_N$

  Replica $d$: $S, d_2, d_3, \ldots, d_N$

  Replica $d$: $S, e_2, e_3, \ldots, e_N$

  where $a_2, b_2, c_2, d_2, e_2$ are the 
neighbors of $S$ in the corresponding replica.

	\item Determine the most common edge among all replicas.
	That is, find which of the five links ($S-a_2$, $S-b_2$, $S-c_2$, $S-d_2$, $S-e_2$)
	appears the most frequently and flag the corresponding replicas. 
	Let's call the most common nearest neighbor city ``$S'$''.
	Suppose there are three replicas containing this link $S-S'$:

  Replica $a$: $S, S', a_3, \ldots, a_N$

  Replica $b$: $S, S', b_3, \ldots, b_N$

  Replica $c$: $S, S', c_3, \ldots, c_N$

	\item Calculate the average position for every $N$ cities. For example, replica $a$ has a long link:
	$S-S'- a_3 - a_4 - \cdots - a_{N-1} - a_N$.
	We can know the distance from a specific city (say $a_4$) to the first city $S$.
	For the relevant replicas (replicas $a$, $b$, and $c$ in this example) 
	calculate the average value of the distance of all cities to the standard city $S$.
\end{enumerate}

\item For replicas that share an identified common edge, attempt to move a random city
to its average position (measured relative to the common edge).
For example, in replicas $a$, $b$, and $c$, we already know the average distance 
of city $a_3$ to $S$.
We compare this value to the distance of all $N$ cities to the standard city,
and we find that the distance of $a_5$ to $S$ is the closest to the average distance 
of city $a_3$ to $S$.
If the total tour length after this change is decreased or if it is increased by less than a specified tolerance,
we rearrange the order as follows: $S, S',a_4, a_5, a_3, a_6, a_7, \ldots, a_N$.
Similarly, we continue to move random cities to their average positions in all relevant replicas for N-2 times.

\end{enumerate}



The geometrical distance coupling $C(A,B)$ between two replicas (candidate tours which share at least one common edge)
is calculated by 
\begin{equation}
\label{cab}
  C(A,B) = \frac{1}{N} \sum_{i=1}^N D_i
\end{equation}
where $N$ is the number of cities in the instance.
$D_i$ represents the difference of the geometrical distance from the $i^\mathrm{th}$ 
city to the standard city in tour $A$ and tour $B$.





\begin{thebibliography}{1}

\bibitem{ref:papadimitriou1998}%
C. H. Papadimitriou and K. Steiglitz, \emph{Combinatorial optimization: algorithms and complexity}, (Courier Dover Publications, 1998).

\bibitem{ref:mezard1987}%
M. M\'ezard, G. Parisi, and M. A. Virasoro,
\emph{Spin glass theory and beyond}, (World scientific, Singapore, 1987), Vol. \textbf{9}.

\bibitem{ref:percus2008}%
A. Percus, et al.,
\emph{Parallel tempering for the traveling salesman problem},
No. LA-UR-08-05100; LA-UR-08-5100. Los Alamos National Laboratory (LANL), (2008).

\bibitem{ref:g222}%
D. E. Goldberg, ``Genetic algorithms in search, optimization, and machine learning.'' 
Addison Wesley (1989).

\bibitem{ref:peter2009}%
P. Ronhovde and Z. Nussinov, Phys. Rev. E \textbf{80}, 016109 (2009).

\bibitem{peter1} P. Ronhovde, S. Chakrabarty, D. Hu, M. Sahu, K. K. Sahu, K. F. Kelton, and N. A. Mauro, and Z. Nussinov,
The European Physical Journal E {\bf 34} 1 (2011).

\bibitem{peter2} P. Ronhovde, S. Chakrabarty, D. Hu, M. Sahu, K. K.  Sahu, K. F. Kelton, N. A. Mauro, and  Z. Nussinov, Scientific reports {\bf 2}, 329  (2012).

\bibitem{dandan1} D. Hu, P. Ronhovde, and Z. Nussinov, Phys. Rev. E {\bf 85}, 016101 (2012).

\bibitem{dandan2} D. Hu, P. Sarder, P. Ronhovde, S. Orthaus, S. Achilefu, and Z. Nussinov, 
Journal of microscopy {\bf 253}, 54 (2014).


\bibitem{dandan3} D. Hu, P. Ronhovde, and Z. Nussinov, Phil. Mag. {\bf 92} (4), 406-445 (2012)




\bibitem{multiplex1}
P. J. Mucha, T. Richardson, K. Macon, M. A. Porter, and J. Onnela,
Science {\bf 328}, 876 (2010).


\bibitem{multiplex2}
J. Gao, S. V. Buldyrev, H. E. Stanley, and S. Havlin,
Nature Physics {\bf 8}, 40-48 (2012).


\bibitem{multiplex3} 
A. Cardillo, J. G\'omez-Garde\~nes, M. Zanin, M. Romance, D. Papo, F. del Pozo, and S. Boccaletti,
Scientific reports {\bf 3}, 1344 (2013).






\bibitem{ref:bianconi2013}%
G. Bianconi, 
Phys. Rev. E {\bf 87}, 062806 (2013).


\bibitem{ref:saad2013}%
R. C. Alamino, J. P. Neirotti, and D. Saad, Phys. Rev. E {\bf 88}, 013313 (2013).


\bibitem{ref:james}
J. Surowiecki, \emph{The Wisdom of Crowds: Why the Many Are Smarter Than the Few and How Collective Wisdom Shapes Business, Economies, Societies and Nations}, (Anchor Books, 2005).


\bibitem{ref:clerc}
M. Clerc, \emph{Particle Swarm Optimization}, (Wiley, 2006).

\bibitem{ref:dorigo1997b}%
M. Dorigo and L. M. Gambardella,
BioSystems \textbf{43}, 73 (1997). 


\bibitem{ref:padberg1991}%
M. Padberg and G. Rinaldi,
SIAM Review \textbf{33}, 60 (1991). 

\bibitem{ref:grotshel1991}%
M. Gr\"otschel and O. Holland,
Mathematical Programming \textbf{51}, 141 (1991).  

\bibitem{ref:johnson1997}%
D. S. Johnson and L. A. McGeoch, ``The traveling salesman problem: 
A case study in local optimization,'' in Local Search
in Combinatorial Optimization (1997), pp. 215-310.

\bibitem{ref:lin1973}%
S. Lin and B. W. Kernighan, Operations Research \textbf{21}, 498 (1973).

\bibitem{ref:Helsgaun2000}%
K. Helsgaun, European J. Operational Research \textbf{126}, 106 (2000).

\bibitem{ref:dorigo1997a}%
M. Dorigo and L. M. Gambardella, Evolutionary Computation,
IEEE Transactions \textbf{1}, 53 (1997).



\bibitem{ref:tsplib}

{\url{http://www.iwr.uni-heidelberg.de/groups/comopt/software/TSPLIB95/tsp/}}

\bibitem{ref:ga2010}
Zakir H. Ahmed, ``Genetic Algorithm for the Traveling Salesman Problem using Sequential Constructive Crossover Operator'', International Journal of Biometrics and Bioinformatics, \textbf{3}, 96 (2010).



\end{thebibliography}
%

\end{document}